\documentclass[twocolumn,trackchanges]{aastex631}

\usepackage{etoolbox}
\pretocmd{\abstractname}{\newpage}{}{}

\usepackage{enumitem}

\newcommand{\kms}{km\,s$^{-1}$}

\shorttitle{{\sl JWST} Survey of Cassiopeia A}
\shortauthors{Milisavljevic, Temim, De Looze, et al.}

\begin{document}

\title{A JWST Survey of the Supernova Remnant Cassiopeia A}

\enlargethispage{5\baselineskip}

\author[0000-0002-0763-3885]{Dan Milisavljevic}
\affiliation{Purdue University, Department of Physics and Astronomy, 525 Northwestern Ave, West Lafayette, IN 47907 }
\affiliation{Integrative Data Science Initiative, Purdue University, West Lafayette, IN 47907, USA}

\author[0000-0001-7380-3144]{Tea Temim}
\affiliation{Princeton University, 4 Ivy Ln, Princeton, NJ 08544, USA}

\author[0000-0001-9419-6355]{Ilse De Looze}
\affiliation{Sterrenkundig Observatorium, Ghent University, Krijgslaan 281 - S9, B-9000 Gent, Belgium}

\author[0000-0003-0913-4120]{Danielle Dickinson}
\affiliation{Purdue University, Department of Physics and Astronomy, 525 Northwestern Ave, West Lafayette, IN 47907 }

\author[0000-0002-3362-7040]{J.\ Martin Laming}
\affiliation{Space Science Division, Code 7684, Naval Research Laboratory, Washington, DC 20375, USA}

\author[0000-0003-3829-2056]{Robert Fesen}
\affiliation{6127 Wilder Lab, Department of Physics and Astronomy, Dartmouth College, Hanover, NH 03755, USA}

\author[0000-0002-7868-1622]{John C.\ Raymond}
\affiliation{Center for Astrophysics $\vert$ Harvard \& Smithsonian, 60 Garden Street, Cambridge, MA 02138, USA}

\author[0000-0001-8403-8548]{Richard G.\ Arendt}
\affiliation{Center for Space Sciences and Technology, University of Maryland, Baltimore County, Baltimore, MD 21250, USA}
\affiliation{Code 665, NASA/GSFC, 8800 Greenbelt Road, Greenbelt, MD 20771, USA}
\affiliation{Center for Research and Exploration in Space Science and Technology, NASA/GSFC, Greenbelt, MD 20771, USA}

\author[0000-0002-4708-4219]{Jacco Vink}
\affiliation{Anton Pannekoek Institute for Astronomy \& GRAPPA, University of Amsterdam, Science Park 904, 1098 XH Amsterdam, The Netherlands}
\affiliation{SRON Netherlands Institute for Space Research, Niels Bohrweg 4, 2333 CA Leiden, the Netherlands }

\author[0000-0003-2317-9747]{Bettina Posselt}
\affiliation{Department of Astrophysics, University of Oxford, Denys Wilkinson Building, Keble Road, Oxford OX1 3RH, UK}
\affiliation{Department of Astronomy \& Astrophysics, Pennsylvania State University, 525 Davey Lab,University Park, PA 16802, USA}

\author[0000-0002-7481-5259]{George G.\ Pavlov}
\affiliation{Department of Astronomy \& Astrophysics, Pennsylvania State University, 525 Davey Lab,University Park, PA 16802, USA}

\author[0000-0003-2238-1572]{Ori D.\ Fox}
\affiliation{Space Telescope Science Institute, 3700 San Martin Drive, Baltimore, MD 21218, USA}

\author[0009-0003-5244-3700]{Ethan Pinarski}
\affiliation{Purdue University, Department of Physics and Astronomy, 525 Northwestern Ave, West Lafayette, IN 47907 }

\author[0000-0001-8073-8731]{Bhagya Subrayan}
\affiliation{Purdue University, Department of Physics and Astronomy, 525 Northwestern Ave, West Lafayette, IN 47907 }

\author[0000-0002-2617-5517]{Judy Schmidt}
\affiliation{Astrophysics Source Code Library, Michigan Technological University, 1400 Townsend Drive, Houghton, MI 49931, USA }

\author[0000-0003-2379-6518]{William P.\ Blair}
\affiliation{The William H. Miller III Department of Physics and Astronomy, Johns Hopkins University, 3400 N. Charles Street, Baltimore, MD 21218, USA}

\author[0000-0002-4410-5387]{Armin Rest}
\affiliation{Space Telescope Science Institute, 3700 San Martin Drive, Baltimore, MD 21218, USA}
\affiliation{The William H. Miller III Department of Physics and Astronomy, Johns Hopkins University, 3400 N. Charles Street, Baltimore, MD 21218, USA}

\author[0000-0002-7507-8115]{Daniel Patnaude}
\affiliation{Center for Astrophysics $\vert$ Harvard \& Smithsonian, 60 Garden Street, Cambridge, MA 02138, USA}

\author[0000-0002-2755-1879]{Bon-Chul Koo}
\affiliation{Department of Physics and Astronomy, Seoul National University, Seoul 08861, Republic of Korea}

\author[0000-0003-3643-839X]{Jeonghee Rho}
\affiliation{SETI Institute, 189 Bernardo Ave., Ste. 200, Mountain View, CA 94043, USA}
\affiliation{Department of Physics and Astronomy, Seoul National University, Gwanak-ro 1, Gwanak-gu, Seoul, 08826, South Korea}

\author[0000-0003-2836-540X]{Salvatore Orlando}
\affiliation{INAF – Osservatorio Astronomico di Palermo, Piazza del Parlamento 1, 90134 Palermo, Italy}

\author[0000-0002-0831-3330]{Hans-Thomas Janka}
\affiliation{Max-Planck-Institut für Astrophysik, Karl-Schwarzschild-Str. 1, 85748, Garching, Germany}

\author{Moira Andrews}
\affiliation{Purdue University, Department of Physics and Astronomy, 525 Northwestern Ave, West Lafayette, IN 47907 }

\author[0000-0002-3875-1171]{Michael J.\ Barlow}
\affiliation{Department of Physics and Astronomy, University College London, Gower Street, London WC1E 6BT, United Kingdom}

\author[0000-0002-3099-5024]{Adam Burrows}
\affiliation{Princeton University, 4 Ivy Ln, Princeton, NJ 08544, USA}

\author{Roger Chevalier}
\affiliation{Department of Astronomy, University of Virginia, P.O. Box 400325, Charlottesville, VA 22904-4325, USA}

\author[0000-0002-0141-7436]{Geoffrey Clayton}
\affiliation{Department of Physics and Astronomy, Louisiana State University, Baton Rouge, LA 70803, USA}

\author[0000-0001-8532-3594]{Claes Fransson}
\affiliation{Department of Astronomy, Stockholm University, The Oskar Klein Centre, AlbaNova, SE-106 91 Stockholm, Sweden}

\author{Christopher Fryer}
\affiliation{Center for Theoretical Astrophysics, Los Alamos National Laboratory, Los Alamos, NM 87545, USA}
\affiliation{Department of Astronomy, The University of Arizona, Tucson, AZ 85721, USA}
\affiliation{Department of Physics and Astronomy, The University of New Mexico, Albuquerque, NM 87131, USA}
\affiliation{Department of Physics, The George Washington University, Washington, DC 20052, USA}

\author[0000-0003-3398-0052]{Haley L.\ Gomez}
\affiliation{Cardiff Hub for Astrophysical Research and Technology (CHART), School of Physics \& Astronomy, Cardiff University, The Parade, Cardiff CF24 3AA, UK}

\author[0000-0002-3036-0184]{Florian Kirchschlager}
\affiliation{Sterrenkundig Observatorium, Ghent University, Krijgslaan 281 - S9, B-9000 Gent, Belgium}

\author[0000-0003-0894-7824]{Jae-Joon Lee}
\affiliation{Korea Astronomy and Space Science Institute, Daejeon 305-348, Republic of Korea}

\author[0000-0002-5529-5593]{Mikako Matsuura}
\affiliation{Cardiff Hub for Astrophysical Research and Technology (CHART), School of Physics \& Astronomy, Cardiff University, The Parade, Cardiff CF24 3AA, UK}

\author{Maria Niculescu-Duvaz}
\affiliation{Department of Physics and Astronomy, University College London, Gower Street, London WC1E 6BT, UK}

\author{Justin D.\ R.\ Pierel}
\affiliation{Space Telescope Science Institute, 3700 San Martin Drive, Baltimore, MD 21218, USA}

\author{Paul P. Plucinsky}
\affiliation{Center for Astrophysics $\vert$ Harvard\ \& Smithsonian, 60 Garden Street, Cambridge, MA 02138, USA}

\author{Felix D.\ Priestley}
\affiliation{Cardiff Hub for Astrophysical Research and Technology (CHART), School of Physics \& Astronomy, Cardiff University, The Parade, Cardiff CF24 3AA, UK}

\author[0000-0002-7352-7845]{Aravind P.\ Ravi}
\affiliation{Department of Physics and Astronomy, University of California, 1 Shields Avenue, Davis, CA 95616-5270, USA}

\author[0000-0003-2138-5192]{Nina S. Sartorio}
\affiliation{Sterrenkundig Observatorium, Ghent University, Krijgslaan 281 - S9, B-9000 Gent, Belgium}

\author{Franziska Schmidt}
\affiliation{Department of Physics and Astronomy, University College London, Gower Street, London WC1E 6BT, United Kingdom}

\author{Melissa Shahbandeh}
\affiliation{Space Telescope Science Institute, 3700 San Martin Drive, Baltimore, MD 21218, USA}

\author[0000-0002-6986-6756]{Patrick Slane}
\affiliation{Center for Astrophysics $\vert$ Harvard \& Smithsonian, 60 Garden Street, Cambridge, MA 02138, USA}

\author[0000-0001-5510-2424]{Nathan Smith}
\affiliation{Steward Observatory, University of Arizona, 933 N. Cherry Ave., Tucson, AZ 85721, USA }

\author{Niharika Sravan}
\affiliation{Department of Physics, Drexel University, Philadelphia, PA 19104, USA}

\author{Kathryn Weil}
\affiliation{Purdue University, Department of Physics and Astronomy, 525 Northwestern Ave, West Lafayette, IN 47907 }

\author[0000-0002-4000-4394]{Roger Wesson}
\affiliation{Cardiff Hub for Astrophysical Research and Technology (CHART), School of Physics \& Astronomy, Cardiff University, The Parade, Cardiff CF24 3AA, UK}

\author[0000-0003-1349-6538]{J.\ Craig Wheeler}
\affiliation{Department of Astronomy, University of Texas at Austin, Austin, TX, USA}

\begin{abstract}

We present initial results from a {\sl JWST} survey of the youngest Galactic core-collapse supernova remnant Cassiopeia~A (Cas~A), made up of NIRCam and MIRI imaging mosaics that map emission from the main shell, interior, and surrounding circumstellar/interstellar material (CSM/ISM). We also present four exploratory positions of MIRI/MRS IFU spectroscopy that sample ejecta, CSM, and associated dust from representative shocked and unshocked regions. Surprising discoveries include: 1) a web-like network of unshocked ejecta filaments resolved to $\sim$ 0.01 pc scales exhibiting an overall morphology consistent with turbulent mixing of cool, low-entropy matter from the progenitor's oxygen layer with hot, high-entropy matter heated by neutrino interactions and radioactivity, 2) a thick sheet of dust-dominated emission from shocked CSM seen in projection toward the remnant's interior pockmarked with small ($\sim 1^{\prime\prime}$) round holes formed by $\la$ 0\farcs1 knots of high-velocity ejecta that have pierced through the CSM and driven expanding tangential shocks, 3) dozens of light echoes with angular sizes between $\sim 0\farcs1$ to $1^{\prime}$ reflecting previously unseen fine-scale structure in the ISM. NIRCam observations place new upper limits on infrared emission ($\la 20$ nJy at 3 $\mu$m) from the neutron star in Cas A's center and tightly constrain scenarios involving a possible fallback disk. These {\sl JWST} survey data and initial findings help address unresolved questions about massive star explosions that have broad implications for the formation and evolution of stellar populations, the metal and dust enrichment of galaxies, and the origin of compact remnant objects. 

\end{abstract}

\keywords{supernovae -- supernova remnants}

\section{Introduction} \label{sec:intro}

Core-collapse supernovae (SNe) are among the most consequential phenomena in the universe, and yet many key questions about their explosion mechanisms and progenitor systems remain unanswered (see, e.g., \citealt{Smartt09} and \citealt{Eldridge13}; \citealt{Janka12} and \citealt{BV21}). This uncertainty in our understanding has broad and important implications: SNe collectively shape the energy balance, chemistry, and structure of galaxies \citep{DS08}; they produce neutron stars (NSs), black holes, and some gamma-ray bursts \citep{Woosley02}; they are a major site of nucleosynthesis \citep{Nomoto13}, dust production \citep{HW70}, and cosmic rays \citep{Blasi13}; and they produce neutrinos \citep{Hirata87,snews21}, and gravitational waves \citep{Murphy09,Andresen17}, that can now be studied with multi-messenger facilities \citep{Szczepanczyk21,Vartanyan23}. 

\begin{figure*}[tp]
\centering

\includegraphics[width=0.63\linewidth]{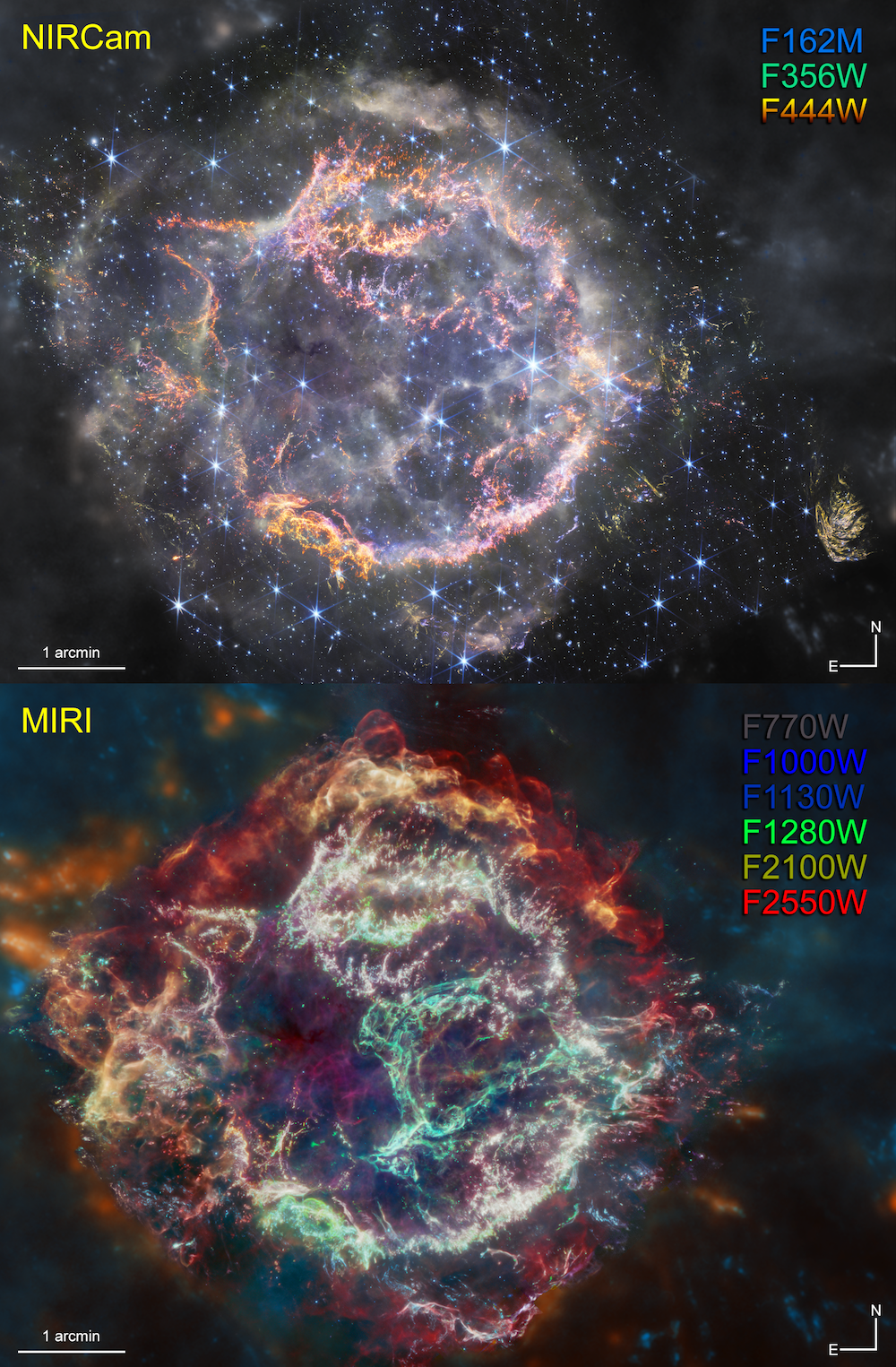}

\caption{Composite images of our NIRCam (top) and MIRI (bottom) mosaics obtained as part of our {\sl JWST} survey of Cas~A. Fields of view have been subtly cropped and minor corrections have been made to compensate for gaps in coverage. Corners not imaged by {\sl JWST} have been filled in with archival {\it Spitzer} data at comparable wavelengths. Mosaics of all individual filters showing the entire fields of view are shown in the Appendix. }

\label{fig:images}
\end{figure*}

The superior resolution and sensitivity to near- and mid-infrared wavelengths made possible with the successful launch of the James Webb Space Telescope ({\sl JWST}; \citealt {Gardner2023}) has opened new pathways to investigate fundamental questions about the nature of SNe via observations of young SN remnants (SNRs).  While recent and upcoming all-sky surveys \citep{Tonry18,Bellm19,Ivezic19} can find tens of thousands of extragalactic SNe, analyses of these unresolved events face unavoidable limitations due to their inability to provide clear and robust three-dimensional kinematic and chemical information.  In contrast, access to many of the specific and detailed properties needed to advance our understanding of massive star explosions can only be obtained through spatially-resolved observations of young ($\lesssim 2000$ yr) Galactic SNRs \citep{Milisavljevic17}.

The near- to mid-infrared spectral region uniquely accesses emission from cool, unshocked SN debris, which enables unique insight into the total mass, relative chemical yield, and kinematic distribution of various components of the SN ejecta \citep{LT20}. It also enables investigations about how much ejecta is transformed into dust and how much of that dust survives passage through the reverse shock \citep{BS07,DA08,WT17,Priestley21}. Finally, the infrared region also avoids much of the problem with foreground extinction towards Galactic remnants, providing new opportunities to constrain processes governing the formation and final fate of compact objects made in SN explosions \citep{DeLuca17}. 
\vspace{-2.5em}

\begin{deluxetable*}{lcccccll}[tp]
  \tablecaption{Image mosaics obtained in the survey \label{tab:observing_log} }
\tablehead{\colhead{Instrument}  &  \colhead{Filter}  & \colhead{$\lambda_p$} & \colhead{BW} & \colhead{PSF}  & \colhead{$t_{\rm exp}$} & \colhead{Date} & \colhead{Sources of strong emission } \\  
                                   & & ($\mu$m) & ($\mu$m) & (\arcsec) & (sec) & \multicolumn{1}{c}{(UT)} & }
                                   \startdata
                                  \hline
				NIRCam & F162M     & 1.626 & 0.168 & 0.055 & 3350 & 2022 Nov 5 & [\ion{Fe}{2}] 1.644; [\ion{Si}{1}]  1.645; synchrotron\\
				               & F356W    &  3.563        &   0.787        &   0.116   & 1675 & 2022 Nov 5 & [\ion{Ca}{4}] 3.207; [\ion{Si}{9}] 3.936; PAHs; synchrotron; dust \\
				               & F444W    &   4.421       &   1.024        &  0.145  & 1675 & 2022 Nov 5 & [\ion{Si}{9}] 3.936; [\ion{Ca}{5}] 4.159; [\ion{Mg}{4}] 4.487; \\
                   & & & & & & & [\ion{Ar}{6}] 4.530;  [\ion{K}{3}] 4.618; CO; synchrotron; dust\\
				 MIRI       & F560W   &  5.6   & 1.2     &        0.207   & 1598      
                                & 2022 Aug 4-5, Oct 22 & [\ion{Mg}{5}]  5.61; dust; synchrotron\\
				                & F770W   & 7.7    & 2.2     & 0.269 & 1598    & 2022 Aug 4-5, Oct 22 & [\ion{Ar}{2}] 6.99; PAHs, dust\\
				                & F1000W & 10.0  & 2.0     & 0.328 & 1598    & 2022 Aug 4-5, Oct 22 &  [\ion{Ar}{3}] 8.991; [\ion{S}{4}] 10.511; dust \\
				                & F1130W & 11.3  & 0.7     & 0.375  & 1598   & 2022 Aug 4-5, Oct 22 & PAHs; dust\\
				                & F1280W & 12.8  & 2.4     & 0.420 & 1598    & 2022 Aug 4-5, Oct 22 & [\ion{Ne}{2}] 12.814; [Ne V] 14.32; dust\\
				                & F1800W   & 18.0  & 3.0     & 0.591 & 1598   & 2022 Aug 4-5, Oct 22 &  [\ion{Fe}{2}] 17.94; [\ion{S}{3}] 18.713; dust;  H$_{2}$\\
				                & F2100W  & 21.0  & 5.0   & 0.674  & 1598   & 2022 Aug 4-5, Oct 22 &  [\ion{S}{3}] 18.713; dust \\
				                & F2550W  & 25.5  & 4.0   & 0.803  & 1598   & 2022 Aug 4-5, Oct 22 &  [\ion{O}{4}] 25.89;  dust\\
				  \enddata
\tablecomments{Values for filter pivot wavelength ($\lambda_p$), bandwidth (BW), and full-width-half-maximum of the point spread function (PSF), have been adopted from {\sl JWST} User Documentation (\url{https://jwst-docs.stsci.edu/}). $t_{\rm exp}$ is the total exposure time for the mosaic. Line identifications guided in part by \citet{Smith09} and \citet{LT20}. }
\end{deluxetable*}

Cassiopeia~A (Cas~A) is arguably the SNR that provides the clearest access to the properties of a core-collapse SN \citep{Koo17}. Cas~A is the youngest Galactic core-collapse SNR known ($\approx 350$ yr; \citealt{Fesen06}); it is among the closest ($3.4^{+0.3}_{-0.1}$ kpc; \citealt{Reed95,Alarie14}); it is the only core-collapse remnant with a secure SN classification from light echo spectroscopy performed from multiple lines of sight (Type IIb; \citealt{Krause08,Rest08,Rest11}); it is one of the best case studies to understand dust formation in SN ejecta and shock-processing of that dust \citep{Rho09,DeLooze17}; and its central X-ray point source is a key object to understanding NS evolution models \citep{Pavlov2009,Gotthelf13,PP22,Shternin23}.  Models for the remnant suggest that the 15--25 M$_{\odot}$ zero-age-main-sequence progenitor star lost the majority of its mass prior to explosion as a $\approx 4$--6\,M$_{\odot}$ star \citep{CO03,HL12,Lee14}, which was likely encouraged through interaction with a binary companion \citep{Young06,Sato20}. Although claims have been made of surviving OB companions in extragalactic SNe IIb \citep{Maund04,Ryder18}, {\sl HST} observations have ruled out this possibility for Cas~A and to date no surviving companion has been located \citep{Kochanek18,Kerzendorf19}.

\begin{figure*}[tp]
\centering

\includegraphics[width=0.85\linewidth]{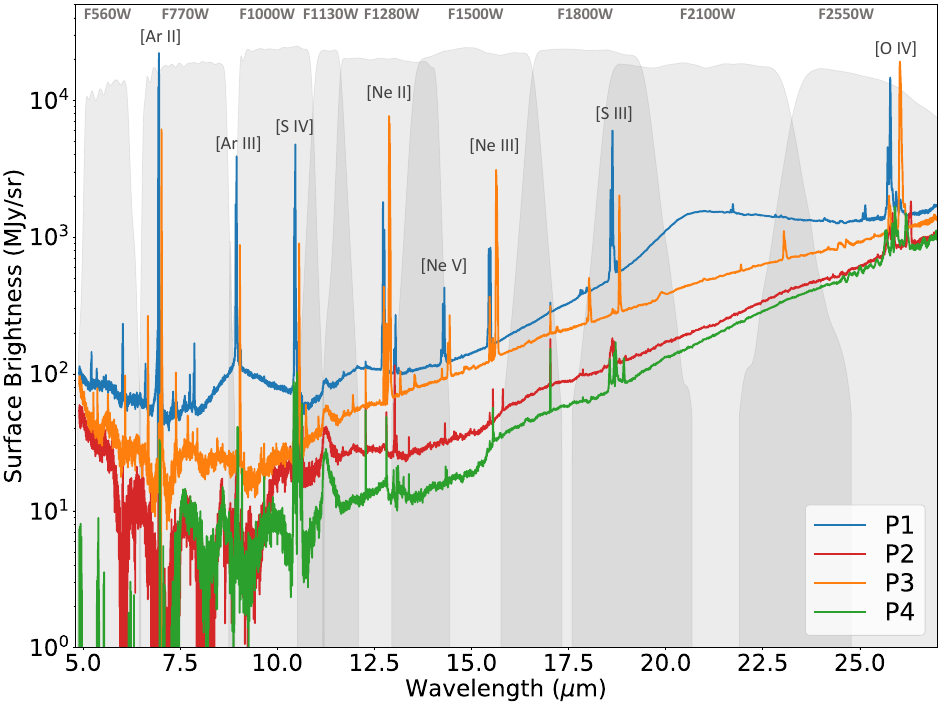}
\caption{MIRI/MRS spectra obtained as part of the survey. P1 and P3 were selected to sample two shocked ejecta knots of different compositions.  P4 represents a `core sample' of unshocked ejecta through the interior of the remnant, and P2 was selected to help diagnose the emission arising in the Green Monster (see Section \ref{sec:mosaics_miri}). All spaxels of the overlapping fields of view across all four channels have been integrated. Positions with respect to Cas A are shown and labeled in Figure~\ref{fig:JWST_Cas_A_Insets}.  Dominant emission lines from the ejecta are labeled. Bandpasses of the MIRI filters are overlaid with throughputs normalized to arbitrary units.  Many of the same emission lines are seen at all four positions, but offsets due to velocity are evident.  Dust emission features (listed in Table~\ref{tab:observing_log}) are seen at all positions with varying intensities and described in more detail in Section~\ref{dust}. 
Note the large dynamic range on the vertical axis. }

\label{fig:spectra}
\end{figure*}

Here we present an overview of a {\sl JWST} reconnaissance of Cas~A made up of near- and mid-infrared imaging mosaics and exploratory spectroscopy. This survey was motivated by outstanding questions about the nature of Cas~A's progenitor system, the explosion dynamics of the original SN, as well as the processes influencing the formation and destruction of dust and molecules. These topics are relevant for broader populations of SNe and their environmental impacts, which in turn have consequences for the formation and evolution of stellar populations \citep{Eldridge08,Smith14ARAA}, the metal enrichment of galaxies \citep{Vogelsberger14,Nelson19}, and the origin of planetary systems \citep{Dwek98,Nittler16}. 

Our NIRCam, MIRI, and MIRI/MRS observations 
are described in  $\S$\ref{sec:observations}, 
followed by $\S$\ref{sec:mosaics} \& $\S$\ref{sec:IFU_spectroscopy} where we present the imaging mosaics and IFU spectra, highlight the data quality, and our major findings with regard to mapping dust and unshocked interior ejecta. We then discuss the serendipitous discovery of a large, bright light echo that resolves the surrounding ISM in $\S$\ref{sec:lightechoes}, and compare our {\sl JWST} data to radio and X-ray observations in $\S$\ref{sec:multiwavelength}. The use of the NIRCam images to constrain possible infrared emission from the surviving NS is discussed in $\S$\ref{sec:neutronstar},
and we review major findings and describe new science opportunities enabled by our survey in $\S8$.  

\section{Observations} \label{sec:observations}

Cas~A was observed with {\sl JWST} in Cycle 1 General Observers (GO) Program 1947 (PI: Milisavljevic). The observations reported here were carried out between August and November 2022, using NIRCam \citep{Rieke2023} and MIRI \citep{Wright2023}.  NIRSpec \citep{nirspec} observations obtained as part of this program that overlap with the MIRI/MRS positions are reported in De Looze et al.\ (in prep.) and \citet{Rho24}. All {\sl JWST} data used in this paper can be found in MAST: \href{https://doi.org/10.17909/szf2-bg42}{10.17909/szf2-bg42}.

The NIRCam observations were obtained on 2022 November 5 using three filters, as shown in Table~\ref{tab:observing_log}. The F162M filter was repeated in the Short Wavelength (SW) camera during both of the Long Wavelength (LW) camera exposures using filters F356W and F444W.  The remnant was covered using a $3 \times 1$ mosaic with 3TIGHT primary dithers, each with 4 subpixel dithers.   The field center is approximately $\alpha (\rm J2000.0)$ = 23:23:23.91, $\delta (\rm J2000.0)$ = +58:48:54.0, with the entire field of view (FOV) spanning approximately $6.3^{\prime} \times 7^{\prime}$ and rotated with position angle 206.8\arcdeg. Some of the resulting mosaics have gaps near the edges of the fields, which depend on the camera. 
The BRIGHT1 readout pattern was used, with seven groups and one integration per exposure, for 12 total dithers leading to a total exposure time of 1675 s.  

\begin{figure*}[tp]
\centering
\includegraphics[width=0.95\linewidth]{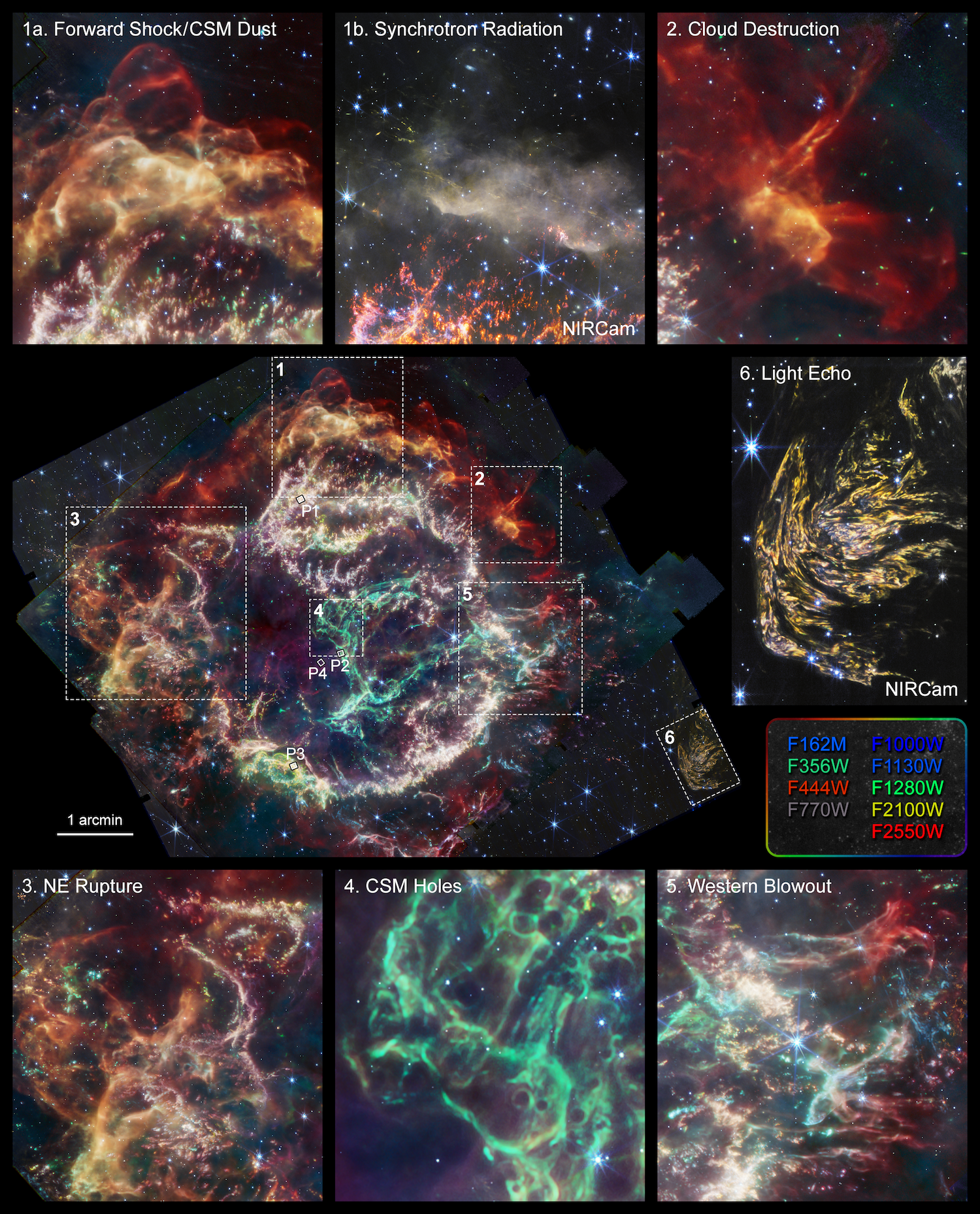}

\caption{Important features of Cas~A identified in our survey and discussed in this paper. The composite image in the center panel combines NIRCam and MIRI filters as indicated. Large boxes outlined with dashed white lines show areas of interest enlarged in the surrounding panels that use the same filters and color scheme, with the exception of panels 1b and 6 that only use NIRCam filters. Small boxes outlined with solid white lines show the positions of the four regions of MIRI/MRS IFU spectroscopy. }

\label{fig:JWST_Cas_A_Insets}
\end{figure*}

Imaging observations with MIRI were first carried out on 2022 August 4-5 using eight filters and are also shown in Table~\ref{tab:observing_log}.  The remnant was covered by a 3$\times$5 mosaic with a 4-point dither pattern. The readout pattern was FASTR1, eight groups and one integration per exposure, with four total dithers leading to exposure times of 88.8 s. An additional location was selected well away from the remnant to sample the background. 

Unfortunately, there was an error with the tiling pattern of the spacecraft, which introduced gaps in the mosaics and a gradual east-to-west drift between the centers of the fields of view of approximately 16$^{\prime\prime}$.  We applied and were approved for a return visit to cover the gaps with three additional MIRI pointings per filter. These return observations were conducted on 2022 October 26 and performed with a position angle approximately 90 degrees to the original visit, which made the edges of the detectors parallel. The overlapping field center of all mosaics is approximately $\alpha$(J2000) = 23$^{\rm h}$23$^{\rm m}$26\fs 97, $\delta$(J2000) = +58$^{\rm o}$$49'$08\farcs 6, with the entire FOV spanning approximately $5.3^{\prime} \times 5.7^{\prime}$ and rotated with position angle 131.5$\degr$. 


The imaging data were processed using the JWST Science Calibration Pipeline \citep{bushouse} version 1.8.4 and the calibration reference data system (CRDS) version 11.16.14, with the CRDS context file $\mathtt{jwst\_1017.pmap}$. Background images were produced from dedicated sky observations and subtracted from the individual images in the $\mathtt{calwebb\_image2}$ pipeline step. We did not subtract background images from the F770W, F1000W, F1130W, and F1280W data due to the variable spatial structure of the sky background at these wavelengths. The level 2 images were astrometrically aligned using the {\sl JWST} Alignment Tool (JHAT) \citep{rest23}.

Mosaic images were then constructed using the default pipeline parameters but with the $\mathtt{tweakreg}$ and $\mathtt{skymatch}$ steps turned off.  Pixel scales of the final mosaics are 0.031$^{\prime\prime}$ per pixel for the NIRCam SW camera images, and 0.063$^{\prime\prime}$ for the NIRCam LW camera images.  The pixel scale of the MIRI mosaics is 0.111$^{\prime\prime}$. Composite images made from the mosaics are shown in Figure~\ref{fig:images}. Mosaics of individual filters showing the entire fields of view are shown in the Appendix.

Spectroscopy was performed between August and November 2022 with MIRI and the Medium Resolution Spectrograph (MRS), which is an integral field unit with a field size that is variable with wavelength ranging from $3.2\arcsec \times 3.7\arcsec$ to $6.6\arcsec \times 7.7\arcsec$.  Three locations were selected after the MIRI imaging observations were available and could be inspected for precise placement with respect to remnant emission: two positions targeted bright ejecta knots in the main shell (P1 and P3), with one targeting newly-identified emission towards the projected center of the remnant (P2). An additional observation P4 was obtained on a location selected using archival {\sl Spitzer} observations targeting unshocked ejecta in the central region \citep{DeLaney10,Isensee10}. For P1, P2, and P3, a four point dither was used with the FASTR1 readout pattern, using 50 groups per exposure, for total exposure times of 555 s. Simultaneous imaging was obtained in the F560W, F770W, and F1500W filters. For P4, the total exposure time was doubled, and simultaneous MIRI imaging was obtained in filters F1130W and F1500W. 

We used version $\mathtt{1.11.0}$ of the $\mathtt{jwst}$ module, which includes the 3 stages of the pipeline processing, and the $\mathtt{jwst\_1141.pmap}$ reference file over the course of the calibration. Stage 1 and stage 2 processing was performed on all four cubes and the designated background with default settings. All cubes ran  through $\mathtt{jwst.residual\_fringe.ResidualFringStep}$ and then stage 3 processing. The designated background was used for the master background step for P1, P2, and P3. The master background step was skipped for P4, since the designated background was obtained 3 months after P4 was observed.

The stage 3 pipeline provides the processed data cubes, as well as a 1D extracted spectrum over the entire field of view. We combined all 12 channel and grating settings. A constant was added to each of the four fully-combined spectra such that the integrated surface brightness over the F770W transmission curve for each spectrum matches the average surface brightness at the corresponding position measured in the F770W image within a region corresponding to the MRS field-of-view at this wavelength.

Below we present and briefly discuss all four MRS spectra, shown in Figure~\ref{fig:spectra}, with locations identified in Figure~\ref{fig:JWST_Cas_A_Insets}. We focus on P2 and P4 within the context of investigating unshocked interior ejecta. De Looze et al.\ (in prep.) provide an in-depth analysis of P2 within the context of interaction between the remnant and the dusty circumstellar environment (see also section \ref{sec:mosaics}), and \citet{Rho24} investigate P1 and P3 within the context of molecule formation and destruction. 

\section{Mosaic Images}
\label{sec:mosaics}

Our mosaic images map thermal and non-thermal near- and mid-infrared emission from Cas~A with unsurpassed depth and sensitivity. Table~\ref{tab:observing_log} provides a complete log of imaging observations along with sources of relatively strong emission in each filter bandpass. Prominent emission features of these imaging mosaics and discoveries of our survey are highlighted in Figure~\ref{fig:JWST_Cas_A_Insets} and discussed below.

\subsection{NIRCam}
\label{sec:mosaics_nircam}

Three filters were selected for the NIRCam mosaics that could distinguish between ejecta, CSM, dust, and the fundamental vibrational mode of CO centered around 4.65 $\mu$m. The F162M filter was selected for its sensitivity to [\ion{Si}{1}] 1.645  $\mu$m and [\ion{Fe}{2}] 1.644 $\mu$m emission present in ejecta \citep{Koo18}, both diffuse and clumped, and He-rich CSM \citep{Koo23}. The F444W filter is sensitive to multiple emission lines of the ejecta, including magnesium and argon, along with CO emission, synchrotron radiation, and faint dust continuum emission. The F356W filter, which largely serves as a continuum reference for the F444W, is sensitive to relatively weaker ejecta lines, including ones from calcium and silicon, along with dust  and synchrotron radiation. 

The strongest emission in the NIRCam mosaic (Figure~\ref{fig:images}) is seen along the main shell of ejecta that has encountered the reverse shock, and represents the remnant's densest material ($n \sim 10^{3-5}$\,cm$^{-3}$) with temperatures between 5000 to 10,000\,K \citep{CK78,HF96,Smith09,Lee17}. As  seen in {\sl HST} data \citep{Fesen01b}, the ejecta are often grouped into large filament complexes of varying scales, and are often increasingly dissipated in directions away from the reverse shock \citep{Morse04}. 

The superb resolution of the NIRCam/F162M (0.055$^{\prime\prime}$) compared to earlier {\sl HST} NIR images places a new constraint on the clump size as small as $<200$ A.U. Furthermore, NIRCam observations detect numerous ejecta knots beyond the main ejecta shell, many of which have not been previously observed either in {\sl HST} optical or ground-based NIR observations \citep{FM16, Koo18}. The knots show different colors in the NIRCam three-color image in Figure~\ref{fig:images}, suggesting their elemental compositions are different. The physical and chemical properties of the outlying ejecta knots will be investigated by Koo et al.\ (in prep.).

Outside of the main shell, the remnant is enveloped in synchrotron emission, seen most strongly around the periphery of the main shell (see representative region enlarged in Figure~\ref{fig:JWST_Cas_A_Insets}, panel 1), but also interior to the main shell  (Figure~\ref{fig:images}). This is associated with the forward shock interacting with surrounding CSM/ISM. Inside the main shell, diffuse emission in F162M is a blend of synchrotron and [\ion{Fe}{2}]+[\ion{Si}{1}] line emission associated with unshocked SN ejecta \citep{Koo18}.  We note that differences in the distribution of interior emission between our F162M image and the long-exposure ground-based image published in \citet{Koo18} are most likely because they removed H-band emission. Additional discussion of the synchrotron radiation is provided in section \ref{sec:multiwavelength}. 

Clumpy N- and He-rich CSM (aka ``quasi-stationary flocculi'' or QSFs; see \citealt{V71}, \citealt{Koo18}, \citealt{Koo20}) appear bright in the F162M image due to their strong [\ion{Fe}{2}] lines. This is evident because all prominent QSFs identified in the 2013 ground-based [\ion{Fe}{2}] 1.64 $\mu$m image by \citet{Koo18} remain observable in the new {\sl JWST} data. {\sl JWST}'s resolution exposes intricate morphologies arising from interactions with the SNR shock. Some QSFs have sharp boundaries with abrupt brightness drops implying the presence of unshocked CSM that has not been processed by the shock yet. Scattered throughout the interior are numerous never-before-seen complete and partial rings approximately 1$^{\prime\prime}$ in size, associated with emission from shocked CSM, that overlap with holes observed in MIRI images (these features are discussed more extensively below in section \ref{sec:mosaics_miri}). 

An unexpected result from our NIRCam mosaics was the discovery of abundant light echoes within close proximity to the remnant, including a particularly large and finely structured one $\sim 4$\arcmin\ southwest of Cas~A's center (Figure~\ref{fig:JWST_Cas_A_Insets}, panel 6). These light echoes are also observed in our MIRI mosaics, in places where the more limited MIRI field of view overlapped with NIRCam. Dozens of light echoes are seen scattered around the periphery and even projected within the main shell, with angular sizes between $\sim 0\farcs1$ to $1^{\prime}$. Light echoes are discussed in more detail in Section~\ref{sec:lightechoes}.

\subsection{MIRI}
\label{sec:mosaics_miri}

For MIRI imaging, eight filters were selected to provide a minimal set required to sample and differentiate line emission associated with ejecta and CSM from dust continuum emission. The bright main shell is made up of reverse-shocked ejecta emitting lines associated with oxygen, sulfur, argon, and neon \citep{Ennis06,DeLaney10}, and an overall much stronger contribution from dust \citep{Rho08}. Clumps of ejecta are seen down to the resolution limit of each image. Dust emission is also present throughout the surrounding CSM that has been heated by the forward shock. Particularly conspicuous is the arc of dust continuum strongest in the F2100W and F2550W filters seen stretching from NW to E outside of the main shell. Bow shocks bright in dust emission are seen in the north, and in the northwest a relatively large cloud of CSM has been overtaken and disrupted by the forward shock (Figure~\ref{fig:JWST_Cas_A_Insets}, panels 1 and 2). 

The strongest emission seen interior to the main shell is concentrated in what we will refer to as the ``Green Monster,'' which extends across the west side of the central region (Figure~\ref{fig:JWST_Cas_A_Insets}, panel 4). The emission is seen faintly in earlier {\sl Spitzer} IRAC and MIPS observations, and the general location overlaps with the characteristic spectra identified by \citet{Arendt14} as the ``South Spot'' that was fit with dust compositions similar to ones found around X-ray Fe ejecta emission. The emission is seen in all filters longward of F1000W, suggesting a dominant dust continuum component. Our {\sl JWST} observations show, for the first time, the detailed structure of the Green Monster, which is concentrated in long filaments among fainter sheets that together are  pockmarked with circular holes $\sim 1^{\prime\prime}$ in radius. Some of these holes are outlined as rings (partial and complete) in the NIRCam mosaics, especially in the F162M filter. 

One possible interpretation of the Green Monster and its unusual pockmarked morphology is that it is associated with unshocked ejecta that have been expanded by small clumps of radioactive material ($^{56}$Ni or $^{44}$Ti).  Cas~A is already known to have been strongly shaped by large-scale Ni-bubbles that expanded into O-, Ar-, and S-rich ejecta, forming large-scale rings as observed in the shocked main shell \citep{Lawrence95,Blondin01-Fe,DeLaney10,Milisavljevic13,Alarie14}, and cavities as observed in the interior unshocked ejecta \citep{MF15}, with radii of 0.5$^{\prime}$ to 1$^{\prime}$. Hence, the holes in the Green Monster could be a much smaller-scale version of this phenomenon, induced by extremely finely clumped radioactive material, expanding within the central interior of the remnant.  

However, in light of low velocity ($\lesssim 100$\,\kms) emission lines of [\ion{Fe}{2}] 1.644 $\mu$m, H Br$\alpha$ 4.05 $\mu$m, [\ion{Ne}{2}] 12.81 $\mu$m, and [\ion{Ne}{3}] 15.56 $\mu$m associated with the small rings (De Looze et al.\ in preparation), along with the X-ray properties of emission in this region that include correlation with morphology observed in the infrared  (\citealt{Vink24}; see also Section~\ref{sec:multiwavelength}), we instead favor that the dominant emission in the Green Monster region is due to circumstellar gas, rich in dust, that has been excited by the forward shock and sculpted by stellar debris. In this scenario, the central location within the main shell is simply a projection effect, and the holes are the result of small knots ($\la$ 0\farcs1) of high-velocity ejecta that have punctured through the CSM and driven expanding tangential shocks (e.g., \citealt{Orlando22}).

Finally, an apparent blowout protruding outward from the main shell is seen in the western portion of the remnant (Figure~\ref{fig:JWST_Cas_A_Insets}, panel 5). This blowout runs due west and is different from the direction of the known southwest outflow (or ``jet'') \citep{Fesen01a,Hwang04}. The morphology of this western blowout shares some resemblance to the NE rupture in the main shell (Figure~\ref{fig:JWST_Cas_A_Insets}, panel 3; see also \citealt{Fesen96Jet}), and the general location is known to host a large population of QSFs \citep{Koo18}. The blowout could be associated with interaction between the remnant and an abrupt density discontinuity in the ISM, not unlike the breakout observed in the NE section of the Galactic remnant CTB\,1 \citep{Fesen97}. Nonetheless, the MIRI emission in this region has conspicuous overlap with i) Si-enriched X-ray emitting ejecta \citep{Vink04} and ii) the projected trajectories of dozens of high-velocity ($\ga 8 \times 10^3$ \kms) S-rich ejecta knots moving radially outward from the center of expansion \citep{FM16}, which together indicate that the blowout, like the NE rupture, may reflect structure imprinted by the explosion dynamics of the original SN \citep{Laming06}.

\section{IFU Spectroscopy}
\label{sec:IFU_spectroscopy}

Our MIRI/MRS spectra only cover four very small fields of view (channel dependent, ranging from $\sim$3.5\arcsec\ to $\sim$7\arcsec\footnote{See the MIRI MRS page in the {\sl JWST} Documentation for details, \url{https://jwst-docs.stsci.edu/jwst-mid-infrared-instrument/miri-observing-modes/miri-medium-resolution-spectroscopy}}), but have far superior spatial and spectra resolution compared to previous low-resolution $5-38$ $\mu$m spectral maps of Cas~A obtained by {\sl Spitzer} with the Infrared Spectrograph (IRS)  \citep{Smith09}. An overview of the spectra of all four positions is shown in Figure~\ref{fig:spectra}.  These MIRI/MRS positions sample a diverse range of emission from the remnant and surrounding environment. P1 and P3 were selected as regions covering different elemental compositions of ejecta and dust excited by the reverse shock in the north and south; both the shape of the dust continuum and the relative intensity of the emission lines vary between the two positions. P2 is coincident with one of the small holes of the Green Monster, but also samples a column of unshocked ejecta and dust through the center of the remnant.  P4 was a ``deep drilling'' IFU position designed to be primarily sensitive to unshocked ejecta, especially iron, through the center of the remnant. Selection of these areas was guided in part by the {\sl Spitzer} IRS spectral maps. Based on the in-flight performance report from \citet{Argyriou2023}, MIRI/MRS wavelength resolution is accurate to 2-27 \kms\ depending on wavelength, and has a spectrophotometric precision of $\approx$5.6\%. 

\subsection{Dust Emission} \label{dust}

Cas A's young age and proximity make it an ideal target for an in-depth investigation of SN dust formation and destruction processes. {\sl Herschel} observations were able to probe the cold dust component in the inner unshocked ejecta for the first time \citep{Barlow10,DeLooze17,Priestley19} and confirmed the high mass of newly formed dust ($0.2-1.0$ M$_{\odot}$), but at a limited spatial resolution of $\sim0.6^{\prime}$. Similarly, the spatial resolution achieved with {\sl Spitzer} IRS spectroscopy (2.5$^{\prime\prime}$ to 6$^{\prime\prime}$) was insufficient to
resolve individual ejecta knots, which made it impossible to constrain the chemical pathways that lead to the formation of specific warm dust species in Cas A.  {\sl Spitzer} IRS did, however, identify the formation of many different dust populations around chemically distinct regions in Cas A \citep{Rho09,Arendt14}. 
{\sl JWST} can associate spatially resolved warm dust knots with different chemical compositions (Ar, Ne, S, O) and has the sensitivity to detect distinct dust emission features 
in those individual knots. The detailed analysis of the MRS dust emission and decomposition of the various dust species that condensed in the ejecta of Cas~A will be presented in a future publication. Here, we summarize the general properties of the dust emission observed with MRS and compare them to previous studies with \textit{Spitzer}.

\begin{figure*}[tp]
\centering

\includegraphics[width=0.95\linewidth]{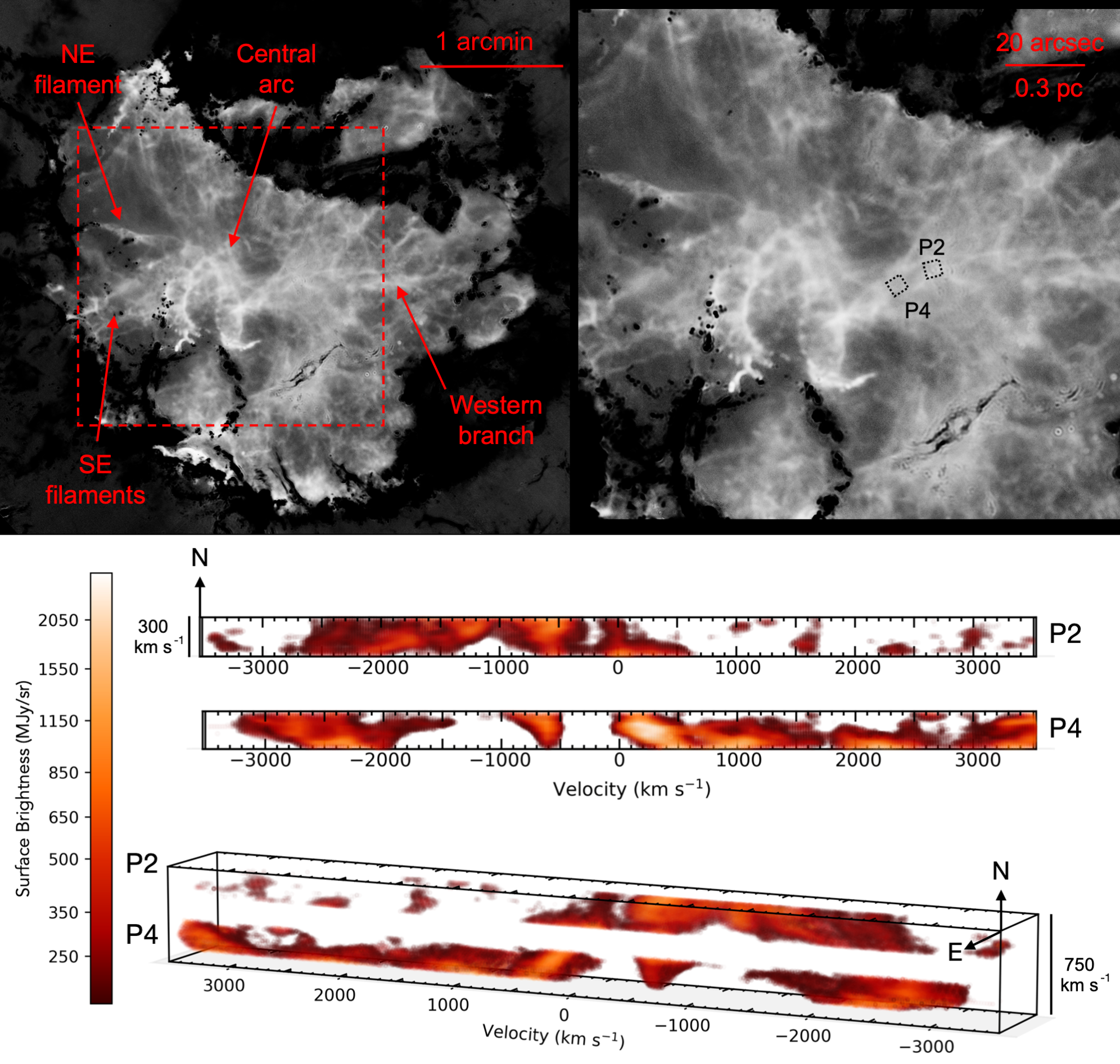}

\caption{Mapping unshocked ejecta of Cas~A. {\it Top}: Image resulting after subtraction of a scaled version of the F2100W filter from F2550W, and masking main shell emission (which appears as black). The emission inside the main shell is mostly due to [\ion{O}{4}] 25.89 $\mu$m from unshocked oxygen material, with potential minor contributions from cold dust, line emission from filaments of shocked oxygen ejecta, and line emission from other shocked/unshocked material including iron. The dashed box indicates the region enlarged in the right panel, and the labeled boxes identify  MIRI/MRS IFU regions. {\it Bottom}: a 3D reconstruction of the MIRI/MRS [\ion{O}{4}] 25.89 $\mu$m emission from P2 and P4 as a function of velocity showing the wide spread of the ejecta in velocity space. Emission  $>3 \sigma$ above the continuum level is represented.}
\label{fig:reddeath}
\end{figure*}

The MIRI/MRS spectra in Figure~\ref{fig:spectra} all show a continuum from warm dust grains and various dust emission features. Position P1, which samples reverse-shocked ejecta with particularly bright argon emission, shows prominent broad dust features at 9, 12, and 21~\micron. \citet{Arendt14} attributed the 9 and 21~\micron\ features to $\rm Mg_{0.7}SiO_{2.7}$ grains characterized by a relatively low magnesium-to-silicon ratio, while \citet{Rho18} attributed the features to SiO$_2$ grains. \citet{Arendt14} argued that the 12~$\micron$ feature could arise from ellipsoidal SiO$_2$ or SiC grains with a continuous distribution of axial ratios, and that the dust with this spectral profile is likely a mixture of silica and Mg silicates with Mg/Si $<$ 5. The MRS spectra show that this 12~\micron\ feature is likely present in all four positions in Cas~A, while the 9 and 21~\micron\ features are only apparent in the argon-rich ejecta. The narrow dust feature at 11.3~$\micron$ that is likely produced by SiC grains \citep[e.g.][]{Jiang05} is also present in all four spectra shown in Figure~\ref{fig:spectra}. We note that the same complex of dust emission features (9, 11.3, 12, and 21~$\micron$) is present in the spectrum of newly condensed SN dust in SNR G54.1+0.3 \citep{Temim10}, which may be evidence that these features all arise from a common grain species or that the multiple grain species that produce the set of features are often found in the same environment. 

Position P3 also samples reverse-shocked ejecta, but it is centered on a region that is neon-bright. This region is characterized by a smooth and featureless dust continuum that is clearly very different from position P1. The continuum emission at positions P2 and P4 that sample the remnant interior and the Green Monster region both show similar shapes with a ``knee'' in the spectrum around $\sim$16~\micron. Interestingly, the shape of the dust spectrum in these interior positions is most similar to the {\sl Spitzer}-observed spectrum arising from Cas~A's CSM component, which \citet{Arendt14} fitted with either $\rm MgFeSiO_{4}$ or $\rm Mg_{2.4}SiO_{4.4}$ as the dominant dust constituent. 

In addition to providing a unique view of dust composition variations, the extensive imaging of Cas A with {\sl JWST} allows the small-scale structures of the freshly condensed ejecta dust to be resolved at unprecedented spatial scales. The high resolution of MIRI images allows ejecta knots to be resolved down to sizes of $1-4\times10^{16}$ cm (or 0.003-0.013 pc), with NIRCam pushing this resolution limit to $6\times10^{15}$ cm (or 0.002\,pc) at 3.56\,$\mu$m. Where {\sl HST} was able to resolve line-emitting ejecta knots down to similar scales, {\sl JWST} provides a complementary view of the dust-emitting structures in the ejecta. Comparison of the specific dust compositions with the distribution and elemental abundances of certain elements on resolved scales can provide us with clues on how dust condensation proceeds. Future work can also use these {\sl JWST} observations of Cas~A to address the predominant question of how much freshly condensed dust will survive passage of the reverse shock. Model estimates for Cas A predict a wide range of survival rates (1-50\%; \citealt{Bocchio16,BC16,Kirchschlager19,Slavin20}). These estimates depend sensitively on their assumed pre-shock grain size distribution and the clump sizes, along with the density contrast between ambient ejecta and dense knots \citep{Kirchschlager23}. {\sl JWST} can resolve the shocked dust emission in ejecta knots down to scales that constrain these parameters with measurements that to date have been impossible to obtain.  

\subsection{Interior unshocked ejecta}
\label{sec:interior}

State-of-the-art core-collapse simulations are having increased success reproducing full-fledged explosions in 3D, and are beginning to evolve explosions from core-collapse to the SNR phase to allow direct comparisons with observations \citep{Hungerford03,Hungerford05,Ellinger2012,Ellinger2013,Wong13,Janka16,Ono20,PM20,Burrows20,Orlando20,Vance2020,Orlando21}. The specific explosion mechanism(s) and appropriate treatment of physics enabling the forward shock to overcome the ram pressure of infalling outer layers remain uncertain.  A neutrino-driven explosion aided by asymmetries is generally favored for Cas~A (\citealt{Hungerford05,Young06,Wong17,Vance2020}), but other explosion mechanisms could participate or even dominate. For example, the two Si-rich wide-angle jet-like outflows of Cas~A in the NE and SW \citep{Fesen01a,Hwang04}, may be the result of magnetohydrodynamic jets \citep{Khokhlov99,MN03,Couch09,Soker18}, or reflect disk accretion of fallback matter by the new-born NS (see section~\ref{sec:neutronstar}). A thermonuclear explosion \citep{KK15}, or quark nova \citep{Ouyed11}, have also been suggested.
  
The relative yields and distributions of elements, including radioactive nuclei freshly synthesized during the SN such as $^{56}$Ni and $^{44}$Ti (and their decay products) that originate from the innermost regions of the star where the explosion initiated, provide valuable constraints for these simulations \citep{Grefenstette14,Wong17,Grefenstette17,fryer23,Sieverding23}.  Hundreds of years after the SN, these nuclei continue to reflect the physical processes dominating the physics of SN engines and probe the degree of explosion asymmetry. 
Particularly important is the mass and distribution of Fe. The X-ray-bright reverse-shocked Fe-rich ejecta of Cas~A (which trace the original $^{56}$Ni distribution) has velocities around and above 4000 \kms\ \citep{Hughes00,DeLaney10} and is associated with a mass of $\approx 0.13$ M$_{\odot}$ \citep{HL12}, which is a fair fraction of the total amount of Fe expected to have been ejected in a neutrino-driven explosion of about $2 \times 10^{51}$\,erg as estimated for Cas~A \citep{Orlando16}.  For this reason, Cas~A has been used as a clear example of how the ejecta asymmetries connected to the neutrino-driven mechanism and the nature of the progenitor as a H-stripped star can strongly influence the final distribution of ejecta by seeding Rayleigh-Taylor instabilities that in turn encourage radial mixing \citep{Wong15,Wong17}. 

Presently, uncertainties in theoretical models (due to both the unknown exact mass and structure of the progenitor and uncertain degrees of freedom in the neutrino physics of supernova calculations) leave it unclear how much unshocked Fe is expected and observations have been unable to derive how much is actually present in Cas~A~\citep[e.g.,][]{Young06,Eriksen09}.  
Multiple lines of evidence support the notion that some Fe must still reside in the center of Cas~A (see, e.g., \citealt{HL12,MF15,Grefenstette17}). However, how much and how it is distributed are critical unknowns. By measuring the iron abundance, we obtain a better understanding of the nature of the central engine.  
Among the best efforts to date to investigate unshocked Fe was a deep 1.64 $\mu$m image of Cas A by \citet{Koo18}, which likely traced interior [Si I] 1.645 $\mu$m line emission as opposed to [\ion{Fe}{2}] 1.644 $\mu$m line, and \citet{LT20} that used {\sl Spitzer} IRS spectra of Cas A. \citet{LT20} estimated the total unshocked ejecta mass to be 0.47$^{+0.47}_{-0.24}$ M$_{\odot}$, which was broadly consistent with previous estimates \citep{HL12,DeLaney14,Arias18}, and with mass fractions of 30\% O, 60\% Si, a few percent S, and traces of Ne and Ar. No Fe was confidently detected in the {\sl Spitzer} data, but possible features at [Fe~VIII] 5.446 and [Fe~V] 20.85 $\mu$m along with other considerations led to an upper limit Fe mass of $< 0.07$ M$_{\odot}$ \citep{LT20}.  

We examined MIRI/MRS spectra of P2 and P4 closely for possible emission features around the [Fe~VIII] 5.446 and [Fe~V] 20.85 $\mu$m emission lines. The only possible candidate is faint emission peaked around 5.39 $\mu$m observed only in the P4 region potentially associated with blue-shifted [Fe~VIII] 5.446 $\mu$m.  A second weaker emission peak is also observed at 5.53 $\mu$m. The distribution of emission around these candidate lines is unlike other ejecta lines that are narrower and better resemble what would be expected from PAHs, though none are documented at these wavelengths. Nonetheless, if this emission originates from truly diffuse, extended gas, then the observed distribution is viable. Following assumptions in \citet{LT20} that emission lines come from density regions where their emission is maximized, we estimate that the putative [\ion{Fe}{8}] emission in the P4 field  of view (approximately $7 \times 10^{-7}$ ergs\,cm$^{-2}$\,s$^{-1}$\,sr$^{-1}$) represents $ \sim 10^{-5}$ M$_{\odot}$ of Fe. If replicated similarly across the remnant, this would imply $\sim 10^{-2}$ M$_{\odot}$ of unshocked Fe material. However, since the candidate emission feature is absent from P2 spectra, clearly the Fe is not distributed uniformly. Future work can investigate the reality of this candidate detection and carefully utilize all available images and spectra to develop the best constraint possible of the total Fe yield from the supernova.

Another potential marker of unshocked ejecta available in the {\sl JWST} range is the relatively bright [\ion{O}{4}] 25.89 $\mu$m line emission (Figure~\ref{fig:spectra}).  Even though ejecta interior to the main shell are unshocked, they are still ionized by the UV and X-ray emission from the surrounding reverse shock, resulting in [\ion{O}{4}]. The F2550W filter is sensitive to [\ion{O}{4}], with additional contribution from dust continuum emission. As seen in Figure~\ref{fig:spectra}, the F2100W filter primarily samples dust continuum emission. By using the F2100W image as a continuum reference, and scaling appropriately to match the general continuum level of F2550W, the resultant of F2550W - F2100W, should primarily arise from [\ion{O}{4}]. 

We present this map of unshocked ejecta in Figure~\ref{fig:reddeath}. A remarkable, asymmetrical, web-like network of filaments resolved to 0.01 pc scales is seen interior to Cas~A's main shell of reverse-shocked material. Many ejecta streams directed outward from the center of expansion are seen, including a thin filament that runs directly towards the center of the NE rupture (``NE filament''), a branch of emission that runs almost due west (``Western branch''), and multiple filaments extending in the SE direction towards the high-energy ejecta plumes of iron observed in the X-ray (``SE filaments'') \citep{Hughes00,DeLaney10,Sato21}. A variety of rings/loops with scales of $\approx 5$\arcsec\ are embedded within these filaments. An especially bright arc of looped filaments is seen east of the main shell center (labeled ``Central arc''). We interpret the significance of this structure in the context of additional multi-wavelength comparisons in section \ref{sec:multiwavelength}.  

We note that portions of our unshocked ejecta map may be contaminated by cold dust embedded in the interior ejecta, as well as emission from filaments of shocked ejecta. However, many arguments strongly support the view that the dominant emission is indeed from unshocked ejecta. For example, the bright ``Central arc'' is consistent with interior structure already mapped in [\ion{S}{3}] $\lambda\lambda$9069, 9531 by \citet{MF15} having a velocity distribution of $-3000$ to $+4000$ \kms. Likewise, the distribution of emission in our unshocked ejecta map strongly correlates with  features interior to the main shell only seen in the [\ion{S}{4}]-sensitive F1000W image.   

\begin{figure*}[tp]
\centering

\includegraphics[width=0.95\linewidth]{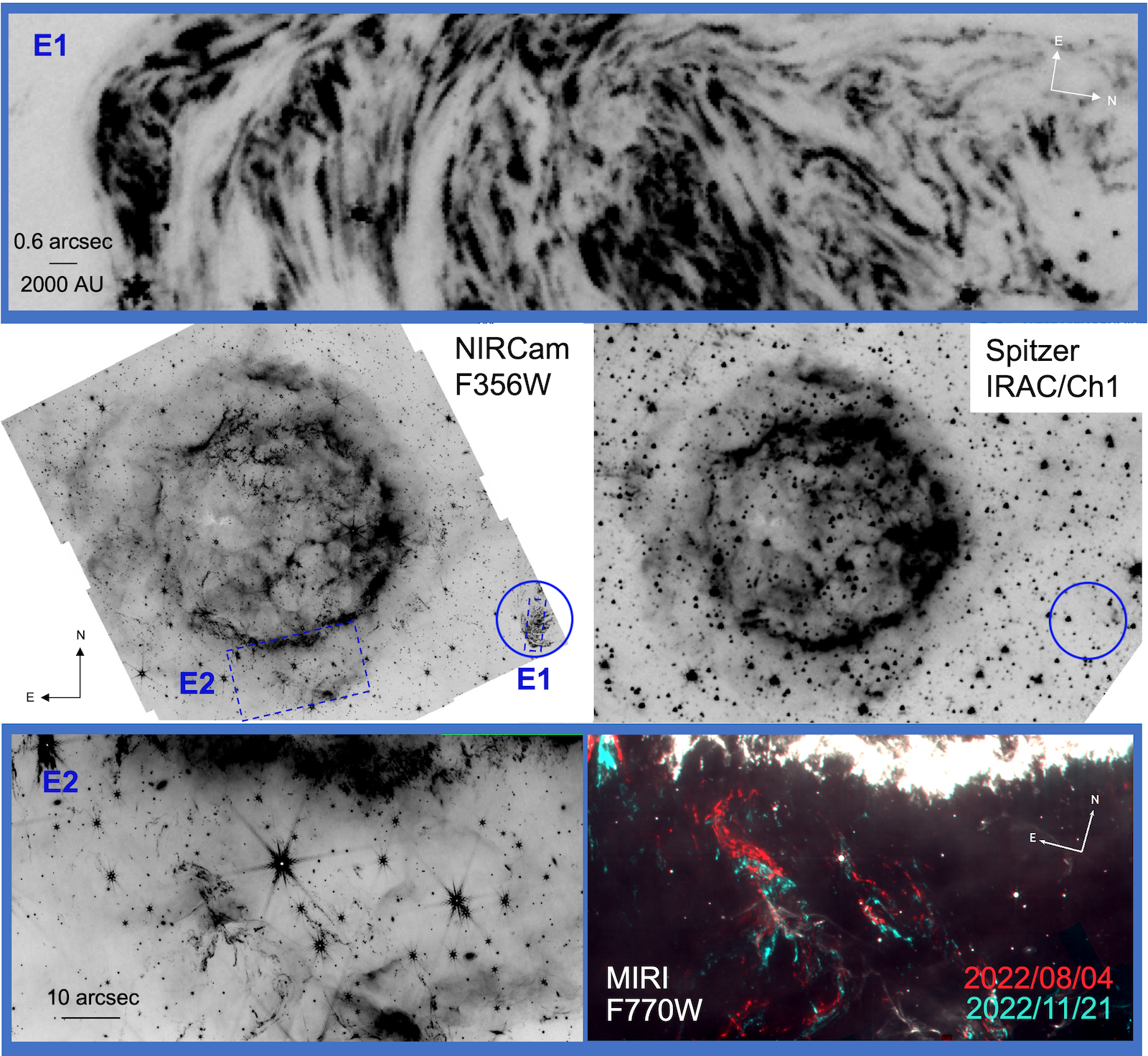}

\caption{Examples of light echoes around Cas~A. Top panel shows an enlarged section of the largest IR echo identified in the middle panels as E1. Comparison of our NIRCam F356W image with archival {\sl Spitzer} IRAC observations (center) shows that the emission was not present in the past. The boxed region E2 highlights an area where multiple epochs of MIRI observations are available and the time variability of the light echoes can be tracked; this region is enlarged in the bottom two panels. In the bottom right panel, sources with negligible proper motion between the 109 days separating the MIRI F770W observations appear white, whereas light echoes appear as red and teal between the first and second observation. The astrometric shift observed in the emission between the two epochs is associated with apparent velocities between 0.3c to 0.6c, which are too large to be motion of SN ejecta.}

\label{fig:lightechoes}
\end{figure*}

The centrally-located P2 and P4 MIRI/MRS spectra provide a complementary probe of the unshocked ejecta through the center of the remnant.  We isolated [\ion{O}{4}] emission in each spaxal by exploring the data cube with the STScI-developed JDAViz tool \citep{jdaviz} to find appropriate wavelength regions to model the continuum.  A 3rd order polynomial was used for the continuum fit, which was then subtracted from the data cube. The cube was cropped by $\sim3$ pixels on each side to eliminate artifacts from the edge of the detector.  Coordinates in right ascension and declination were converted into velocity using the scaling factor of 0.022$^{\prime\prime}$ per \kms\ \citep{Milisavljevic13}, with respect to the center of expansion \citep{Thorstensen01}. The data were linearly interpolated and resampled with a resolution of 22 \kms\ in all three dimensions, which is comparable to the original spatial resolution. 

The continuum-subtracted, cropped, interpolated [\ion{O}{4}] emission from P2 and P4 as a function of velocity is shown in the bottom of Figure \ref{fig:reddeath} from two viewing angles. Previous observations by {\sl Spitzer} using the IRS presented by \citet{Isensee10} with spatial resolution of $2.5^{\prime\prime}$ and spectral resolution of 0.05 $\mu$m indicated sheet-like structures and filaments, including holes within the sheets at the scale of 30\arcsec. Our data, which localize ejecta velocities with uncertainty $\la$160 \kms\ for any 3D location of the [\ion{O}{4}] emission and improve over {\sl Spitzer} observations by approximately an order of magnitude, reveal an even more intricate distribution with filaments seen at the scale of 1\arcsec. The velocities range over $\pm$3000 \kms, indicative of ejecta and not CSM.  We interpret the MIRI/MRS spectra as further confirmation that the interior network of filaments seen in the top panel in Figure~\ref{fig:reddeath} traces unshocked ejecta accurately.

The [\ion{O}{4}] emission is accompanied by fainter emission in the [\ion{S}{4}] 10.521 $\mu$m, [\ion{S}{3}] 18.676 $\mu$m and [\ion{Ar}{3}] 21.823 $\mu$m lines. Their spatial structure within the IFU field and their velocity structure are nearly identical to those of [\ion{O}{4}], implying that these elements are thoroughly mixed down to the scale of the {\sl JWST} resolution.  In photoionization equilibrium in Cas~A, these ions exist at densities in the range of 1 - 20 amu/cc and temperatures from about 20,000\,K down to 600 K \citep{LT20}.  This gas fills around 20\% of the volume.  If higher density gas were present, the [\ion{Ar}{2}] 6.9842 $\mu$m line would be visible, but only a very narrow line from the background or photoionized CSM is seen.  There is no emission from Ne or the [\ion{Fe}{2}] and [\ion{Fe}{3}] lines in the spectra of the unshocked ejecta, whereas they are seen in the shocked ejecta at P1 and P3. Together this indicates that substantial abundance variations occur on larger scales.

\begin{figure*}[tp]
\centering

\includegraphics[width=\linewidth]{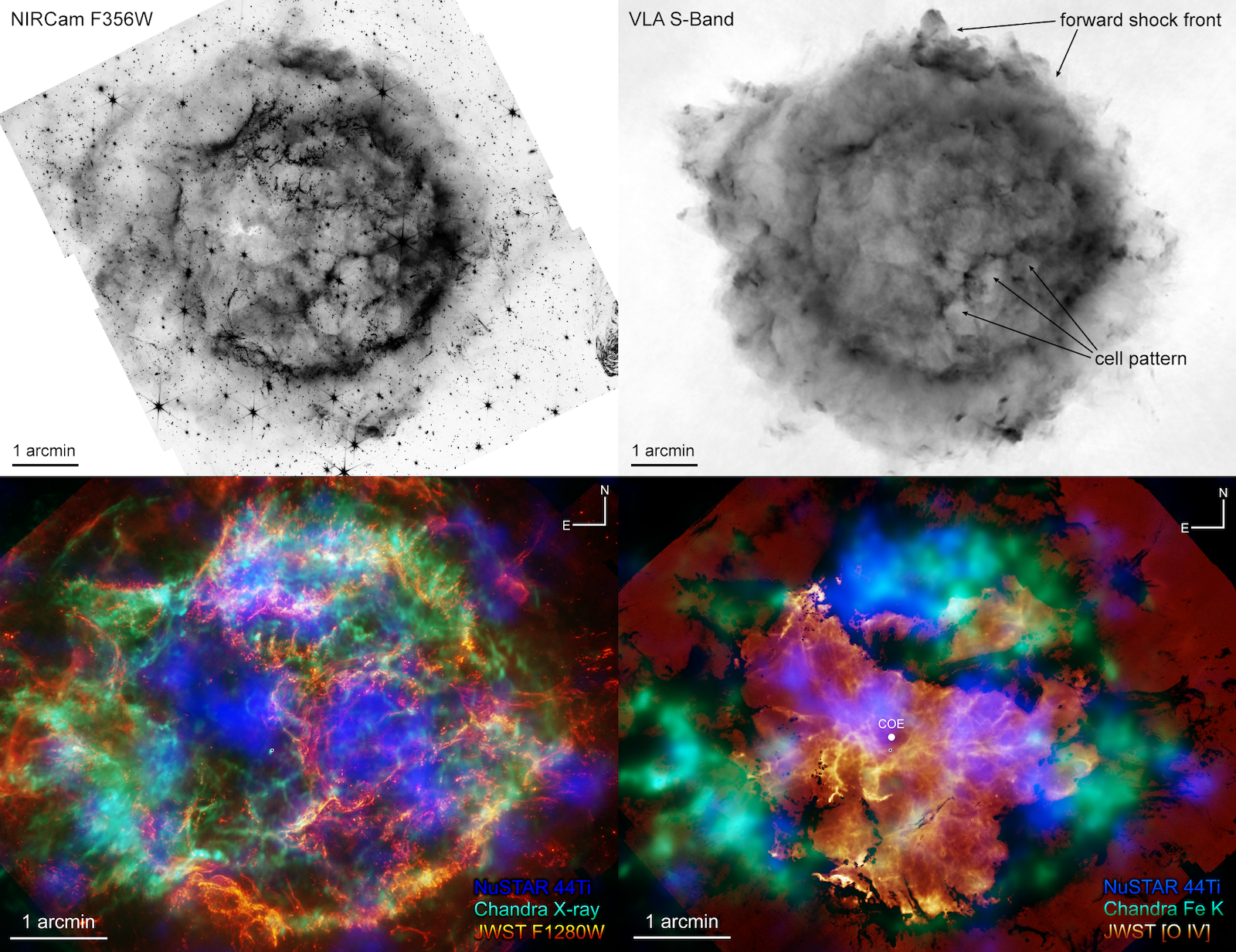}

\caption{Top: The NIRCam F356W image compared to a VLA S-band image from 2012 (courtesy of T.\ DeLaney). Bottom: Left panel shows a composite image made using the {\sl NuSTAR} $^{44}$Ti map \citep{Grefenstette14}, the {\sl Chandra} 1\,Msec X-ray image of Cas~A \citep{Hwang04}, and our MIRI F1280W image. The location of the NS \citep{Fesen06-CCO} is marked with a black circle. The right panel shows a composite image combining our unshocked ejecta map (see Section~\ref{sec:interior}), with the {\sl NuSTAR} $^{44}$Ti map,  and a {\sl Chandra} Fe--K map (energy $\sim 6.7$ keV). The locations of the center of expansion (COE; \citealt{Thorstensen01}) and the NS are marked with white and black circles, respectively.}

\label{fig:mutliwavelength}
\end{figure*}

\section{Light echoes} \label{sec:lightechoes}

Transient luminous sources such as novae and SNe can produce echoes of their outbursts by way of reflected light from dusty regions in the surrounding  ISM \citep{vdb65,Rest05,DA08}. Light echoes have been recorded around Cas A and spectra of them have been used to classify the explosion type \citep{Krause08}, measure explosion asymmetry by observing the original supernova from different lines of sight \citep{Rest11}, and probe the structure of the surrounding ISM itself \citep{Kim08,Vogt12,Besel12}. Generally, emission at $\gtrsim 3$ $\mu$m is likely to be reradiated thermal emission from dust, whereas the shorter wavelength emission is more likely to be scattered light of the transient source itself. Light echoes can be recognized by their structure (more sharply defined than ISM features), spectral energy distribution (warmer than typical ISM), and most clearly by their dramatic brightness changes (when data are available at multiple epochs).

Echoes of various brightness are found across the field of our {\sl JWST} observations. Figure 5 highlights two prominent regions of echoes. The region labeled E2 shows typical bright echoes, which appear as bright clumps embedded in fainter more filamentary structures. The additional {\sl JWST} observations discussed in section~\ref{sec:observations} were performed 109 days following the initial observations, providing an opportunity to search for variations of light echo features surrounding portions of Cas A. The lower right panel in Figure~\ref{fig:lightechoes} shows the dramatic changes seen in the MIRI F770W images between these two epochs. These can be confidently identified as light echoes since the proper motion of emission, if interpreted as the same clumps of material from one epoch to the next, would imply unrealistic velocities between 0.3c to 0.6c. The fastest moving ejecta in Cas A is traveling at 0.05c in the outermost portion of the northeast jet \citep{FM16}.

Even more surprising was the appearance of a particularly large and bright IR echo in the corner of the NIRCam field (Figure ~\ref{fig:JWST_Cas_A_Insets}, panel 6), enlarged in Figure~\ref{fig:lightechoes}. Comparison with archival {\sl Spitzer} observations confirms that this emission was not present previously, and detection across multiple filters ensures this is a real feature. Recent {\sl HST} imaging in December 2022 (GO 17210; PI: Fesen) also weakly detected this new light echo, further verifying its reality. An echo this close to Cas A ($\approx 4^{\prime}$ from the center of expansion) means the reflecting surface is almost directly behind the remnant ($\sim 170$ lt-yr; see further \citealt{DA08}). In this orientation, the echoing ellipsoid provides a tomographic slice of that portion of the ISM, nearly in the plane of the sky. 

The incredible structure in this echo, showing many small isolated features and concentric arclets, along with its overall well defined outer edge, indicates a very complex ISM cloud surface.  The thinnest strands remain unresolved in width even in our NIRCam F162M image. To our knowledge, this is the best view yet of ISM structure at the smallest of scales possible: 0.1$^{\prime\prime}$ translates to $\approx 350$ A.U., or $\sim 2$ light-days, which is consistent with the echoes changing completely on time scales of months \citep{DA08}, and implies that changes on time scales of hours are possible. The F162M was observed in parallel with both the F356W and F444W observations, which occurred 19.3 hours apart in the initial mapping. Comparison of the F162M images of the large IR echo at these epochs does not reveal any very short term changes in the echo brightness. This suggests that there were no major changes in the illuminating SN emission on this time scale and/or that the echoing structures are $\gtrsim 19$ lt-hr $\simeq 140$ A.U. in size. However, F162M is not the band that is  most sensitive to the echo emission. 

It is worthwhile to note that \citet{Krause05} had speculated that some of the echoes may have been produced more recently ($\sim1952$) by an unseen or highly directional burst from the NS. Although \citet{Kim08} and \citet{DA08} discounted this possibility based on geometry and energetics required to fit the IR spectra, some suspicion lingers due to the locations, morphology, and apparent motion of some echoes, which give the visual impression that they lie within the circumstellar environment and thus must be of recent origin. Optical (i.e., reflected) spectra of these echoes in the immediate vicinity of Cas A would definitively determine if they are the product of the Type IIb SN explosion \citep{Krause08}, or a more recent NS outburst of some kind.

\section{Multi-wavelength comparisons}
\label{sec:multiwavelength}

Cas~A exhibits complex and overlapping sources of thermal and non-thermal radiation, arising from ejecta, CSM, dust, molecules, and particle acceleration. Comparing our {\sl JWST} infrared observations with {\sl Chandra X-ray Observatory} and Jansky Very Large Array radio telescope data make it possible to disentangle these many sources of emission.

Cas A is one of the brightest radio sources on the sky \citep{Reber44}, having a total flux density of $\sim2000$ Jy in the 1-2 GHz band \citep{Baars77}. A relatively circular region with radius $\sim 100^{\prime\prime}$ is associated with the reverse shock propagating into the expanding ejecta. Weaker radio emission associated with synchrotron radiation \citep{GS65} is seen at some azimuths extending out $\sim 150^{\prime\prime}$ to the location of the main forward shock, traveling at 4200 – 5200 \kms\ \citep{DR03,PF09}. 

Infrared emission shortward of about 5 $\mu$m is known to be largely synchrotron emission from electrons accelerated in shocked regions \citep{Rho03,Jones03}. Earlier {\sl Spitzer} observations with IRAC at 3.6 $\mu$m were observed to resemble radio synchrotron images \citep{Ennis06}. 
A more recent study involving radio and {\sl Spitzer} IRAC data revealed evidence for spectral flattening from a spectral index of $\alpha\approx 0.77$ in the radio to $\alpha\approx 0.55$ for the radio to infrared comparison, in particular for those regions close to shock regions \citep{domcek21}. However, in the southeast region  the synchrotron spectra are relatively steep $\alpha\approx 0.67$, perhaps indicating the onset of synchrotron cooling. Synchrotron cooling should affect the synchrotron spectrum for frequencies $\nu \gtrsim 7.2\times 10^{13}~B_{\rm mG}^{-3}t_{100}^{-2}$~Hz, with $B_{\rm mG}$ being the magnetic-field strength in milliGauss (mG) and $t_{100}$ the age of the shocked plasma in units of 100 yr. The F356W filter corresponds to $\nu\approx 8.4\times 10^{13}$~Hz. This implies that signatures of synchrotron cooling can be present for the plasma containing relativistic electrons that have been accelerated early on ($t_{100}\approx 3$), and/or for $B\gtrsim 1$~mG. For comparison,  estimates for the mean magnetic-field strength in Cas A are $\sim 0.5$--1~mG \citep[e.g.][]{rosenberg70,vink03a,uchiyama08}.

In Figure~\ref{fig:mutliwavelength} (upper panels), we show the NIRCam F356W image alongside an S-band VLA image of Cas~A. The increase in sensitivity and spatial resolution of {\sl JWST} over {\sl Spitzer} is clear in {\sl JWST}'s ability to map the faint, extended synchrotron radiation component. At least at a qualitative level, these images also reveal patches of low infrared synchrotron surface brightness in the southeastern region, in agreement with \citet{domcek21}. However, a more quantitative analysis accounting for the variable background level is required to provide solid evidence for synchrotron cooling effects in the southeastern region, and perhaps elsewhere.

Other striking morphological features, most likely associated with synchrotron radiation, are the conspicuous cell-like patterns concentrated in the central southwest region (Figure~\ref{fig:mutliwavelength}). Although these features appear to have a radio counterpart, the contrast between the patterns and more diffuse synchrotron emission is lower. A preliminary explanation for the infrared pattern is that it is the consequence of regions observed edge-on, perhaps associated with blowout regions on the near- or far-sides of Cas~A, or with protrusions in the reverse shock \citep{Blondin01-RT,Mandal23}. If the infrared synchrotron emission from these regions is affected by synchrotron cooling, then the shells surrounding these blowout regions are narrower in the infrared than in the radio, providing stronger limb-brightening effects. Portions of the cell pattern align with X-ray synchrotron filaments, for which synchrotron cooling effects are even more prominent. 

In Figure~\ref{fig:mutliwavelength} (lower left panel), we show a composite image using 0.3-10 keV {\sl Chandra} broadband observations \citep{HL12}, a {\sl NuSTAR}  $^{44}$Ti map \citep{Grefenstette14}, and our MIRI F1280W image. A gap in the $^{44}$Ti emission that runs north-south is approximately where the brightest ridge of the Green Monster is located. Because we favor the interpretation that the bulk emission of the Green Monster is associated with CSM on the front side of the remnant (towards us), the anti-correlation with $^{44}$Ti (which is redshifted overall; \citealt{Grefenstette17}) should be coincidental. Absorption of $^{44}$Ti emission ($\approx 70$\,keV) from the Green Monster seems unlikely, since a CSM column density of $\sim 10^{27}$\,cm$^{-2}$ would be needed to reduce the flux by a factor of 10. Nonetheless, clear correlations in emission are seen between {\sl Chandra} and our {\sl JWST} data, suggesting a shared origin. De Looze et al.\ (in preparation) and \citet{Vink24} investigate this X-ray--IR relationship in detail and attribute it to interaction between the forward shock and relatively dense CSM.

Another composite image is shown in Figure~\ref{fig:mutliwavelength} (lower right) that compares our unshocked ejecta map (section \ref{sec:interior}; Figure~\ref{fig:reddeath}) with a {\sl Chandra} Fe--K map (energy $\sim 6.7$ keV; tracing where $^{56}$Ni was produced), and the {\sl NuSTAR}  $^{44}$Ti map. As noted earlier, the SE filaments and Western branch of the unshocked ejecta connect with Fe-rich locations. On the other hand, no strong relationship is seen between [\ion{O}{4}] and $^{44}$Ti, except that  $^{44}$Ti emission is located where [\ion{O}{4}] is weak. Together, these comparisons support the view that the fine structures in the interior volume are most likely rooted in turbulent mixing of cool, low-entropy matter from the progenitor's oxygen layer with hot, neutrino and radioactively heated high-entropy matter.

\section{Neutron Star}
\label{sec:neutronstar}

\begin{figure*}[tp]
\centering
\includegraphics[width=\linewidth]{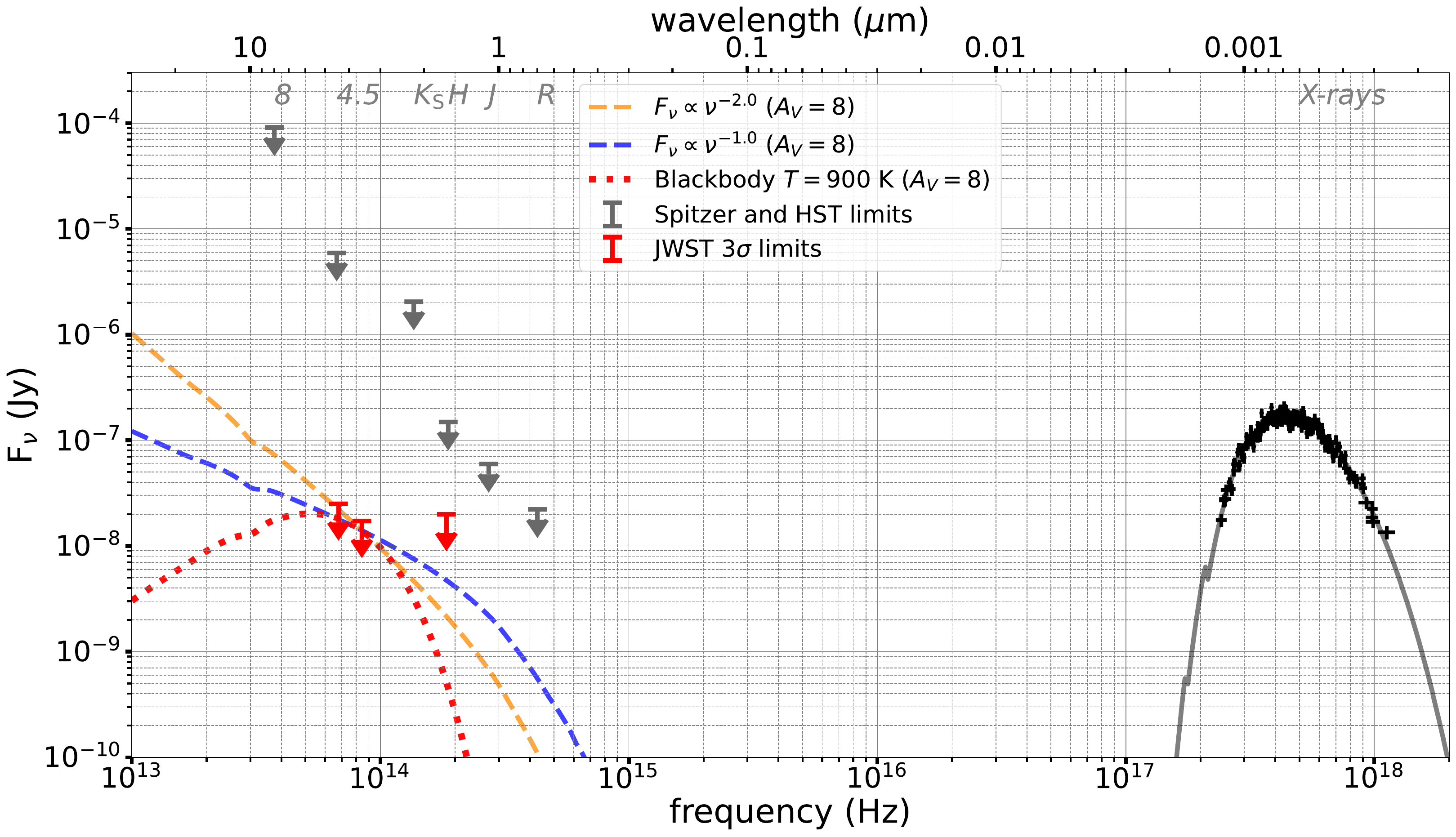}
\caption{The {\sl JWST} $3\sigma$ upper limits in the near-infrared (NIRCam F162M, F356W, F444W) at the location of the Cas\,A CCO and the CCO's X-ray spectrum detected with {\sl Chandra} ACIS. Power law spectra consistent with {\sl JWST} upper limits and an example blackbody spectrum with a temperature of 900\,K are plotted for an extinction of $A_V=8$, using the extinction curve by \citet{Gordon2023}. The {\sl Chandra} ACIS X-ray data are from \citet{Posselt2018}, the {\sl HST} optical and NIR limits are from \citet{FPS06}, and the {\sl Spitzer} limits are from \citet{Wang2007}. }

\label{fig:CCOmw}
\end{figure*}

First-light images of Cas A taken by {\sl Chandra} revealed a central X-ray point source \citep{Tananbaum99,pavlov2000}, that is presumably its remnant NS.  This NS is part of the family of enigmatic {\it central compact objects} (CCOs), which are young, exhibit X-ray emission that is steady and predominantly thermal, and lack surrounding pulsar wind nebulae and counterparts at other wavelengths \citep{Pavlov04,HG10,DeLuca17}.  CCOs have relatively low inferred magnetic fields, $B\sim (3$--$10) \times 10^{10}$\,G for the three CCOs where it was estimated (i.e., two orders of magnitude lower than that of a typical young NS), and X-ray luminosities larger than their spin-down power \citep{Gotthelf13}.

Deep {\sl HST} NICMOS images of the Cas~A center in  NIR filters \citep{FPS06} did not detect the CCO. The {\sl JWST} NIRCam images enable a much deeper look at the CCO of Cas~A, which is regarded as a key object to understanding NS evolution. The Cas~A CCO is the youngest of the dozen currently known CCOs, as well as the youngest of all the known NSs for which we can directly study the surface emission. It is particularly interesting because it has possibly shown a noticeable cooling during the 20+ years of {\sl Chandra} observations \citep[][and references therein; but see \citealt{PP22} and \citealt{Alford2023}]{Shternin23}.

Because ongoing accretion can be excluded for CCOs \citep{Pavlov04}, their thermal X-rays are likely emitted from the NS surface layers heated by the heat transfer from the cooling NS interiors.
The low magnetic fields of CCOs can be explained by two scenarios:
(1) these NSs are born with weak magnetic fields (``anti-magnetars'')
or (2) the normal ($\sim 10^{12}$\,G) or strong ($\gtrsim 10^{14}$\,G) magnetic field of a newborn NS has been buried in its crust by ``fallback'' debris after the SN explosion \citep{Vigano12}. Such fallback could also form a disk surrounding the NS. Comprehensive searches at X-ray and radio wavelengths have failed to find evolutionary descendants of these young CCOs \citep{Gotthelf13}.
Disk emission in the IR would strongly support the re-emergence of magnetic field scenario and explain why it is so difficult to identify CCO descendants.

Non-detections outside the X-ray range are usually attributed to a lack of pulsar activity because of the low magnetic field or to high extinction \citep{Pavlov04,HG10,DeLuca17}.  Our NIRCam images offer the possibility to peer deeply enough through high extinction at wavelengths other than X-rays to test hypotheses about the CCO's nature. To this end, we conducted forced photometry at the coordinates of the CCO ($\alpha$(J2000) = 23$^{\rm h}$23$^{\rm m}$27\fs 943, $\delta$(J2000) = +58$^{\rm o}$$48'$42\farcs 5; \citealt{FPS06}), on the three NIRCam filter images  using the \textit{space\textunderscore phot}\footnote{See \url{https://app.soos.io/research/packages/Python/-/space-phot}.} package. The photometry was measured using point-spread function (PSF) photometry on the level 2 NIRCam ``CAL'’ images, which preserves the PSF structure compared to level 3 resampled images. The NIRCam pixel area map for the corresponding SW and LW modules must also be applied to each exposure to correct for pixel area variations across the images. Here we use the PSF models provided by WebbPSF (version 1.2.1)\footnote{\url{https://www.stsci.edu/jwst/science-planning/proposal-planning-toolbox/psf-simulation-tool}} to represent the PSF, which have been updated to better match the observed PSF in each filter, and take into account temporal and spatial variation across the detector. We remove a constant background in the fitting region, which is small enough that a constant background is sufficient. The background is estimated as the median value in an annulus centered at the position with an inner radius of 5 pixels and outer radius 7 pixels. \textit{space\textunderscore phot} simultaneously constrains the common flux in all CALs at the forced photometry position, within a $5\times5$ pixel square. The final measured flux is the integral of each full fitted PSF model, which includes a correction to the infinite aperture flux. These total fluxes, which are in units of MJy/sr, are converted to AB magnitudes using the native pixel scale of each image.

No sources were detected confidently, and we estimate 3$\sigma$ AB magnitude upper limits of 28.15, 28.31 and 27.90, corresponding to 20.0, 17.3 and 25.2 nJy for the F162M, F356W, and F444W images, respectively. There are three faint source candidates within an arcsecond of the CCO coordinates with S/N $\approx 2-3$ in the F162M image. However, these candidate sources are consistent with detector noise patterns seen elsewhere, and are not present in other filters, so we judge these to not be credible CCO candidates. We performed the same procedure on a modified version of the F162M image where the so-called $1/f$ noise had been removed by custom processing the stage 2 pipeline output and rerunning the stage 3 pipeline, but this exercise did not yield a deeper limit. 

Thus, although our limits are much deeper than those found by \citet{FPS06}, as seen in Figure~\ref{fig:CCOmw}, the CCO remains undetected. We should note that even these deep limits are much higher than the expected thermal emission from the NS with $T_{\rm eff}=2\times 10^6$ K (e.g., the predicted flux density at 3\,$\mu$m is $\sim 1.4 \times 10^{-13}$\,Jy). However, the measured {\sl JWST} limits can be used to constrain some possible scenarios in which IR emission could be expected.

Firstly, we can constrain the area of any circular disk with a blackbody temperature  similar to the one inferred from the {\sl Spitzer} detection of the magnetar 4U\,0142+61 \citep {Wang06}. Such a disk with temperature $\sim 900$\,K must have an area smaller than 
$5.5\times 10^{20} (\cos i)^{-1}$ cm$^2$ (i.e., $R_{\rm out} < 0.2 (\cos i)^{-1/2} R_\odot$), where $i$ is the disk inclination, and $R_{\rm out}$ is the outer radius of the disk.

The temperature of a realistic fallback disk is expected to decrease outwards. If the decrease can be approximated by a power law, $T(r) =T_{\rm in} (r/R_{\rm in})^{-\beta}$, where 
$T_{\rm in}$ is the temperature at the inner disk radius $R_{\rm in}$, then the spectrum of the disk emission is $f_\nu \propto \nu^{3 - (2/\beta)}$ at $kT_{\rm out} \lesssim h\nu\lesssim kT_{\rm in}$, where $T_{\rm out} = T_{\rm in}(R_{\rm in}/R_{\rm out})^\beta$ is the temperature at the outer edge of the disk. For instance, the limiting power-law spectra $f_\nu\propto \nu^{-1}$ and $\nu^{-2}$, shown in Figure \ref{fig:CCOmw}, correspond to $\beta=0.5$ and 0.4, respectively.

If the disk is heated by X-rays produced by the NS, one can expect its IR flux to be proportional to the X-ray flux. For the Cas A's CCO, the upper limit on the IR to X-ray flux ratio is $F_{\rm F356W}/F_{\rm 0.6-6\,keV} < 5 \times 10^{-6}$. This limit is substantially lower than  $F_{\rm F356W}/F_{\rm 0.6-6\,keV} \sim 4\times 10^{-5}$ for the magnetar 4U\,0142+61, or $F_{\rm F160W}/F_{X}= 3\times 10^{-4}$ for
the extended NIR emission around the much older ($\sim 0.5$\,Myr) X-ray thermal isolated neutron star RX\,J0806.4$-$4123 \citep{Posseltetal2018}.

A power law spectrum is also expected for non-thermal emission from the NS or from a pulsar wind nebula.  In the case of the Cas\,A CCO, the non-thermal X-ray emission may be faint and overpowered by the strong thermal emission from the NS surface, but nonthermal emission could in principle be detectable at IR wavelengths, where the thermal NS emission is undetectable. The new {\sl JWST} limits, however, indicate that any non-thermal NS emission must be very faint.   

\section{Conclusions}

We have presented results from an extensive {\sl JWST} survey of the SN remnant Cas~A made up of NIRCam + MIRI imaging mosaics and exploratory MIRI/IFU spectroscopy. Our observations provide the most detailed and comprehensive mapping ever of its ejecta,  CSM/ISM, and associated dust and molecules at NIR and MIR wavelengths. Significant findings and implications made from this survey include the following:

\begin{enumerate}[leftmargin=*]

    \item We have uncovered a web-like network of unshocked ejecta filaments seen in [\ion{O}{4}] emission and resolved to $\sim$ 0.01 pc scales reflecting turbulent mixing processes and hydrodynamical instabilities that occurred shortly after core collapse of the progenitor star. We weakly detect emission possibly associated with [\ion{Fe}{8}] 5.446 $\mu$m that had been  predicted by \citet{LT20}, and estimate upwards of $\sim 10^{-2}$ M$_{\odot}$ of diffuse, unshocked Fe material. Many aspects of this distribution are relevant for core-collapse SN simulations, including a clear connection seen between the unshocked ejecta filaments and {\sl Chandra} Fe--K emission. Correlations with {\sl NuSTAR} are less clear but worth further investigation. This finding is also relevant for attempts to simulate light curves and spectra of unresolved extragalactic SNe, which must make assumptions regarding the fragmentation of the ejecta due to hydrodynamical instabilities. Ni-bubble expansion reduces the effective optical depth of non-radioactive material \citep{Dessart18,DA19}, and inappropriate treatment could mean that the ejecta masses for the majority of stripped-envelope SNe are systematically underestimated \citep{EF22,Ergon23}.
    
    \item A large structure called the ``Green Monster,'' interior to the main shell and bright in MIRI images, is resolved for the first time. We conclude that this is most likely associated with a thick sheet (or sheets) of dust-dominated emission from shocked CSM seen in projection toward the remnant's interior. The structure is pockmarked with small ($\sim 1^{\prime\prime}$) round holes that have most likely been formed by high-velocity N- and He-rich ejecta knots that have pierced through the CSM and driven expanding tangential shocks. Support for this conclusion, including detailed analysis of MIRI/MRS and NIRSpec spectroscopy and correlated X-ray emission observed with {\sl Chandra}, is provided in De Looze et al.\ (in preparation) and \citet{Vink24}. The discovery provides an exciting opportunity to probe mass loss of the progenitor system some $10^{4-5}$ years prior to explosion.

    \item Dozens of light echoes with angular sizes between $\sim 0\farcs1$ to $1^{\prime}$ reflecting previously unseen structure in the ISM are found throughout our NIRCam and MIRI images. Although light echoes had been expected in the vicinity of Cas A, the brightness, size, and especially the complexity of an echo seen in the SW corner of our NIRCam mosaics is quite startling. Its fine-scale structure suggests relatively poor mixing, and raises questions of how it came to be this way. Is the dust not mixed uniformly into the ISM gas, or is the gas also structured this way? Was this once some sort of concentric shells of outflow from a star that have been buffeted and twisted by turbulence in the ISM? Are magnetic fields involved? If this structure is typical of the large ISM clouds, do we see the imprint of this on older SNRs in some way? A more detailed investigation of the large SW echo and other light echoes may start to answer these questions. 

    \item We do not detect the neutron star in the center of Cas~A. Our NIRCam observations place new upper limits on the CCO's infrared emission ($\la 20$ nJy at 3 $\mu$m) and tightly constrain scenarios involving a possible fallback disk.
    
\end{enumerate}

Additional epochs of {\sl JWST} observations would make it possible to improve characterization of the kinematic and elemental properties of the ejecta via proper motion measurements, while also permitting further study of nearby light echoes. Imaging mosaics with wider fields of view (reaching 5\arcmin\ from the center of expansion versus the $\approx$3\arcmin\ reach of our existing data) would increase opportunities for these measurements dramatically, and cover the NE/SW outflows to more accurately constrain their total mass, composition, and relationship to the original SN \citep{FM16}. Additional spectroscopy, including widefield NIRSpec mirco-shutter assembly observations across the entire SNR and targeted MIRI/MRS observations of regions interior to the main shell (especially ones that overlap with $^{44}$Ti emission), would be invaluable to improve estimates of chemical abundances and place the deepest limits possible for the total Fe yield. 

\bigskip


We thank the referee for a very helpful reading of the Letter and suggestions that significantly improved its content and presentation. This work is based on observations made with the NASA/ESA/CSA James Webb Space Telescope. The data were obtained from the Mikulski Archive for Space Telescopes at the Space Telescope Science Institute, which is operated by the Association of Universities for Research in Astronomy, Inc., under NASA contract NAS 5-03127 for {\sl JWST}. These observations are associated with program \#1947. Support for program \#1947 was provided by NASA through a grant from the Space Telescope Science Institute, which is operated by the Association of Universities for Research in Astronomy, Inc., under NASA contract NAS 5-03127.

This research was supported by the Munich Institute for Astro-, Particle and BioPhysics (MIAPbP) which is funded by the Deutsche Forschungsgemeinschaft (DFG, German Research Foundation) under Germany´s Excellence Strategy – EXC-2094 – 390783311. We are grateful for the Lorentz Center for hosting the workshop ``Supernova Remnants in Complex Environments'' where some ideas used in this manuscript were developed. Tracey DeLaney provided {\sl Spitzer} IRS and VLA S-band data that were inspected as part of this paper. Brian Grefenstette provided the NuSTAR $^{44}$Ti map used in this paper. Una Hwang provided the {\sl Chandra} 1\,Msec image used in this paper. D.M.\ is grateful for conversations with Luc Dessart, who contributed ideas used in this paper. D.M.\ acknowledges invaluable support from Mark Linvill and Chris Orr who help manage computational resources used to complete this research.

D.M.\ acknowledges NSF support from grants PHY-2209451 and AST-2206532. T.T.\ acknowledges support from the NSF grant 2205314 and the JWST grant JWST-GO-01947.031. J.M.L.\ was supported by JWST grant JWST-GO-01947.023 and by basic research funds of the Office of Naval Research.
I.D.L.\ acknowledges funding from the Belgian Science Policy Office (BELSPO) through the PRODEX project “JWST/MIRI Science exploitation” (C4000142239). I.D.L., F.K., and N.S.S.\ have received funding from the European Research Council (ERC) under the European Union’s Horizon 2020 research and innovation programme DustOrigin (ERC-2019-StG-851622). M.J.B.\ acknowledges support from the ERC grant SNDUST ERC-2015-AdG-694520.
M.M.\ and R.W.\ acknowledge support from STFC Consolidated grant (ST/W000830/1). S.O.\ acknowledges support from PRIN MUR 2022 (20224MNC5A). B.-C.K.\ acknowledges support from the Basic Science Research Program through the NRF of Korea funded by the Ministry of Science, ICT and Future Planning (RS-2023-00277370). Work by R.G.A.\ was supported by NASA under award number 80GSFC21M0002. The research of J.C.W.\ is supported by NSF AST-1813825. H.-T.J.\ is grateful for support from the German Research Foundation (DFG) through the Collaborative Research Centre ``Neutrinos and Dark Matter in Astro- and Particle Physics (NDM),'' Grant No.\ SFB-1258-283604770, and under Germany's Excellence Strategy through the Cluster of Excellence ORIGINS EXC-2094-390783311.

\facilities{JWST(MIRI), JWST(NIRCam)}

\appendix

\renewcommand\thefigure{A\arabic{figure}}    
\setcounter{figure}{0}   

\begin{figure}[b]
\centering

\includegraphics[width=0.83\linewidth]{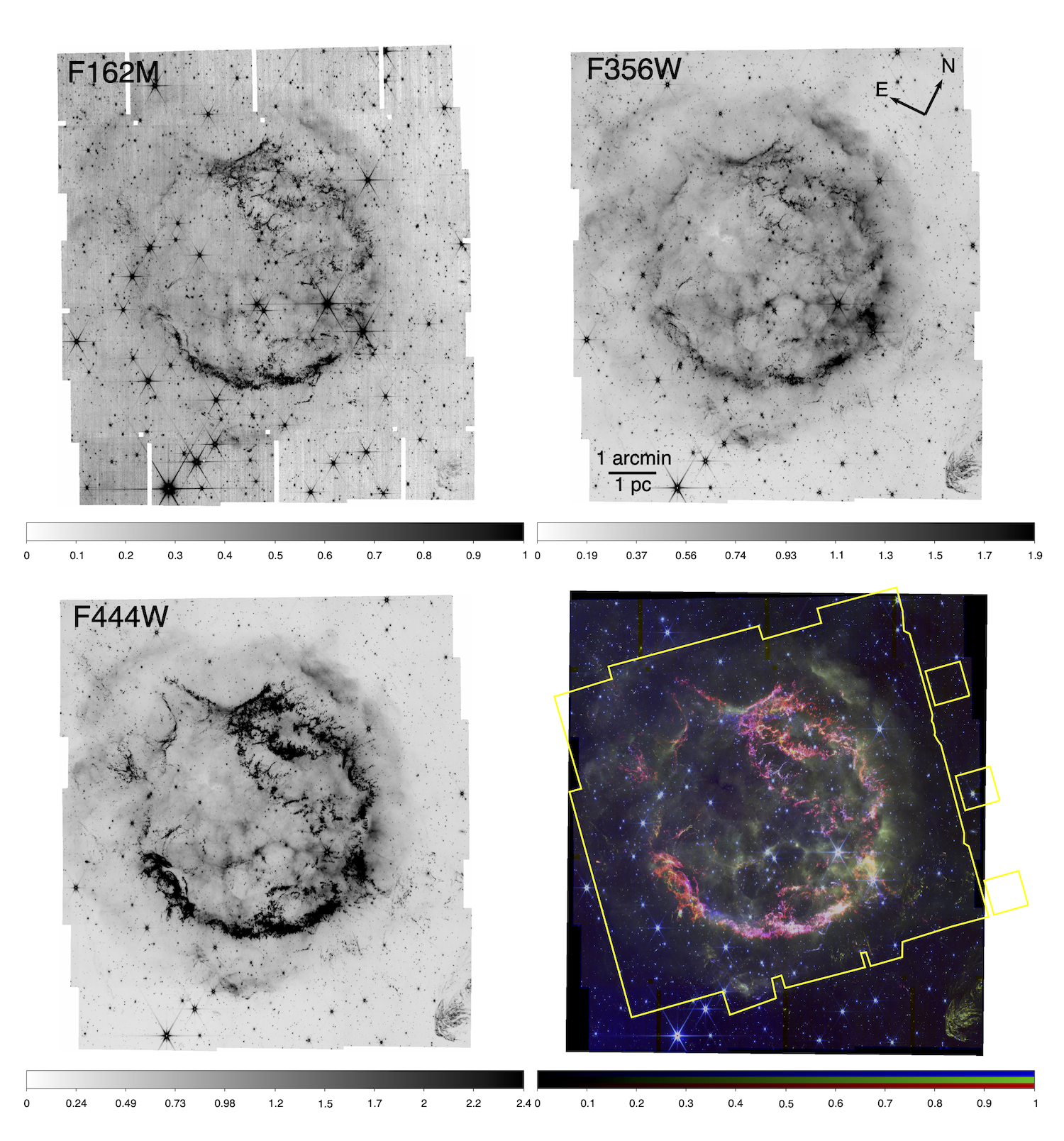}

\caption{Three NIRCam imaging mosaics obtained as part of our Cas A survey. The F162M filter was used as the SW channel for both the F356W and F444W observations, so the exposure time is twice as long.  Gaps are seen in the F162M mosaic because of the dithering 3-point pattern that prioritized the LW filters and introduced gaps in the SW channel detectors. The images are with the same angular scale and orientation, and the intensity scale is in units of MJy sr$^{-1}$. In the color version at lower right, the blue, green and red show the images in increasing wavelength, respectively. The approximate footprint of the MIRI mosaics is shown.}

\label{fig:appendix-nircam}
\end{figure}

\begin{figure*}[tp]
\centering

\includegraphics[width=\linewidth]{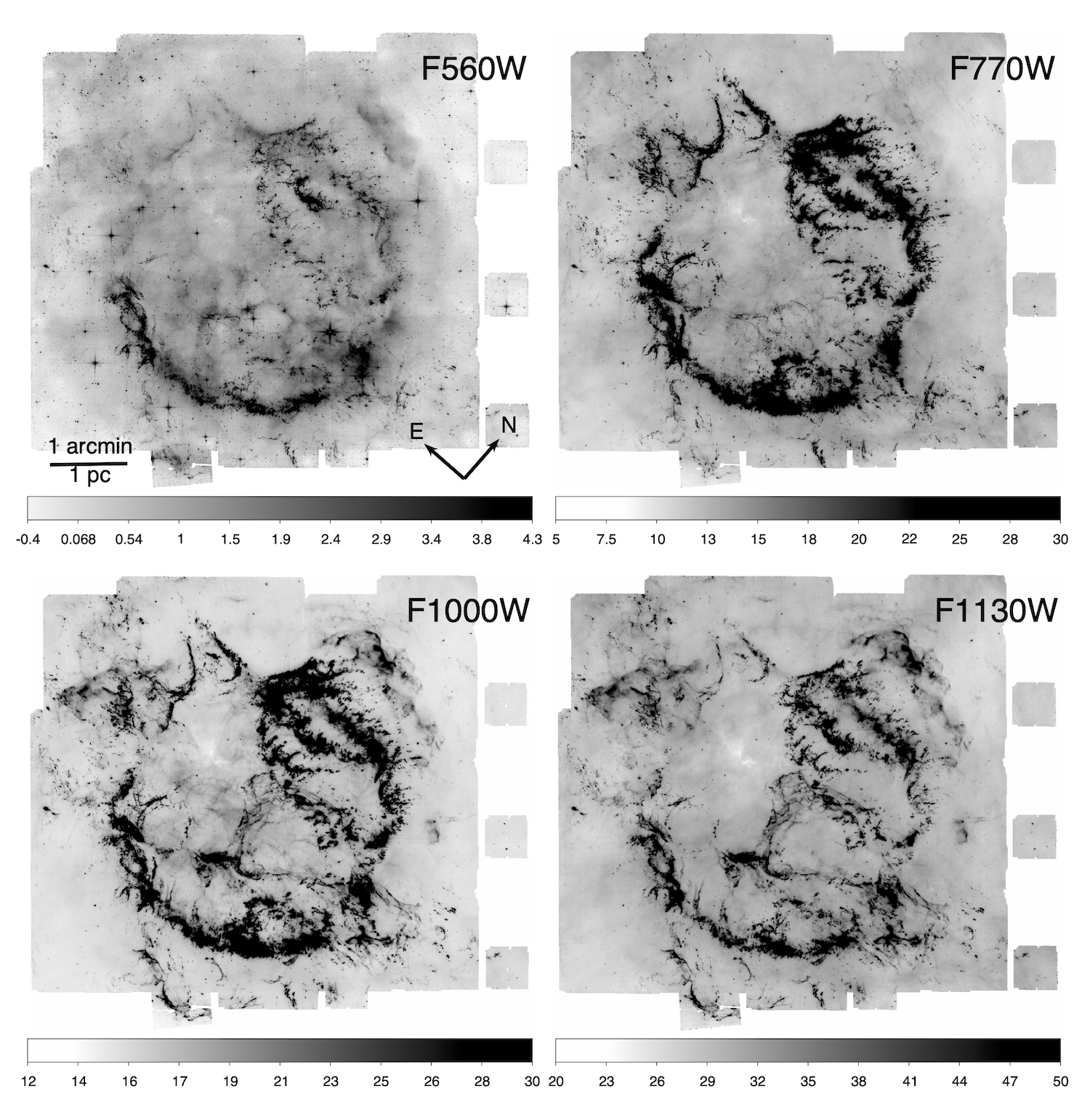}

\caption{The four shorter wavelength MIRI imaging mosaics obtained as part of the survey. The full spatial coverage (including Lyot coronagraph) is shown. The common angular scale and orientation is shown in the upper left panel. Color bars depict the intensity scale that has units of MJy\,sr$^{-1}$.}

\label{fig:appendix-miri-short}

\end{figure*}

\begin{figure*}[tp]
\centering

\includegraphics[width=\linewidth]{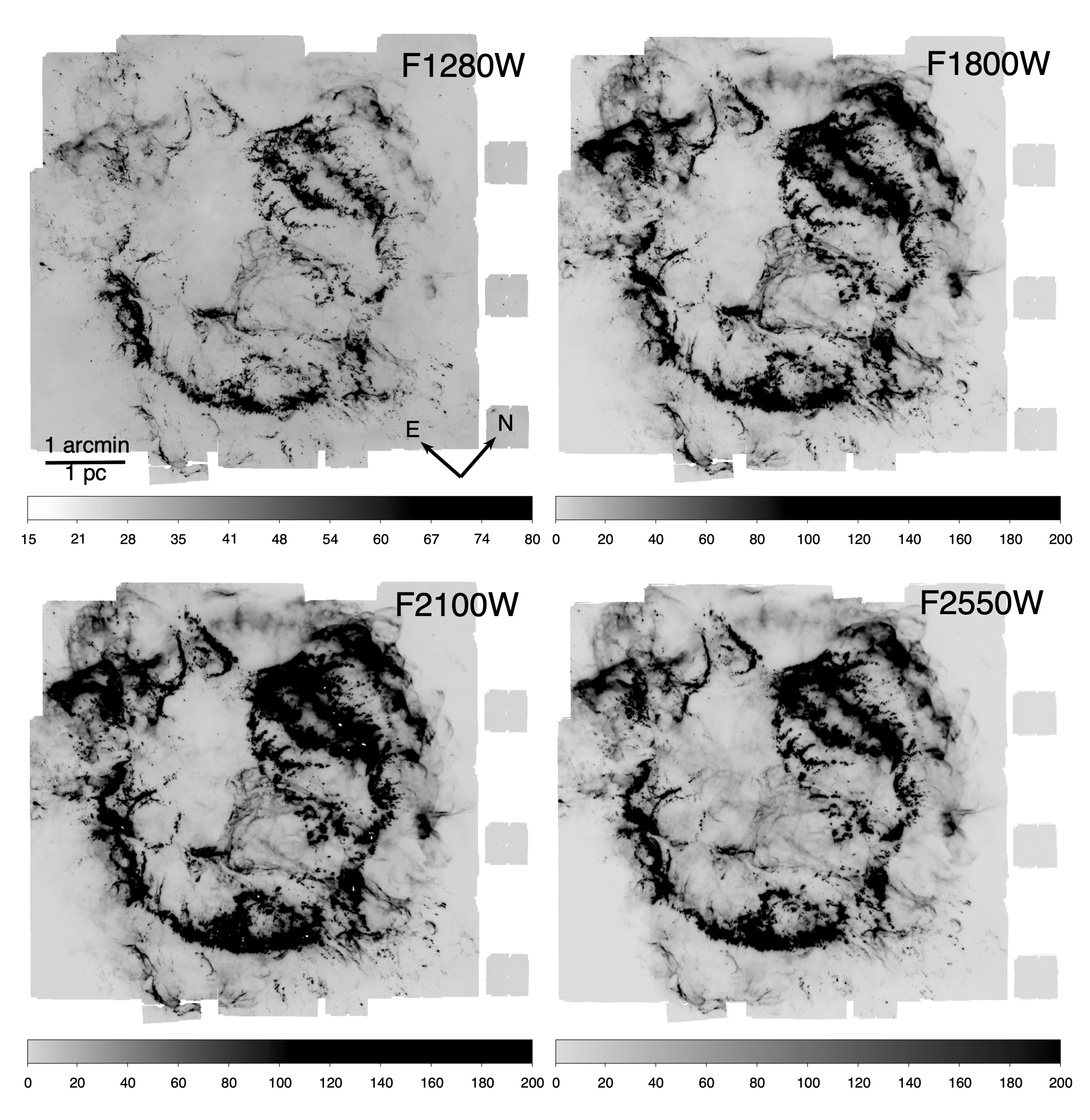}

\caption{Same as Figure~\ref{fig:appendix-miri-short}, except showing the four longer wavelength MIRI imaging mosaics.}
\label{fig:appendix-miri-long}
\end{figure*}

The color composite images shown in the body of this paper are helpful for showing the relative differences between multiple filters of data, but they can also hide details that are present within the individual data sets, especially for our MIRI imaging. As indicated in Figure~\ref{fig:spectra}, some bandpasses are dominated by continuum emission and others contain both continuum and line emission. In Figures \ref{fig:appendix-nircam}, \ref{fig:appendix-miri-short}, and \ref{fig:appendix-miri-long} of this appendix, we show images of each filter of the NIRCam and MIRI imaging mosaics individually to allow these details to be assessed on a filter-by-filter basis.

\end{document}